\documentclass{my_iopjournal}

\usepackage{amsmath}
\usepackage{amssymb}
\usepackage{multicol}
\usepackage{ragged2e}
\usepackage{dcolumn}
\usepackage{bm}
\usepackage[utf8]{inputenc}
\usepackage[T1]{fontenc}
\usepackage{mathptmx}
\usepackage{float}
\usepackage{lipsum}
\usepackage{etoolbox}
\usepackage[separate-uncertainty=true]{siunitx}
\usepackage{booktabs}
\usepackage{xfrac}
\usepackage[colorlinks=true, allcolors=blue]{hyperref}
\usepackage{threeparttable}
\usepackage{makecell}
\usepackage{titlesec}
\usepackage[backend=bibtex, style=numeric-comp, sorting=none, maxnames=99]{biblatex}
\DeclareSIUnit{\dBm}{dBm}

\titlelabel{\thetitle.\enskip}

\begin{document}

\title{Thin-film nanostructuring of NbTi superconducting resonators for enhanced temperature sensitivity in cryogenic thermometry}

\author{
André Chatel\textsuperscript{\textcolor{blue}{1},\textcolor{blue}{2},\textcolor{blue}{*}}\orcid{0000-0003-3016-0104},
Roberto Russo\textsuperscript{\textcolor{blue}{1},\textcolor{blue}{2}}\orcid{0000-0002-6489-888X}, 
Seyed Alireza Hashemi\textsuperscript{\textcolor{blue}{1}},
Pol Torres-Vila\textsuperscript{\textcolor{blue}{1}}\orcid{0009-0009-2891-1694},
Jürgen Brugger\textsuperscript{\textcolor{blue}{1}}\orcid{0000-0002-7710-5930}, 
Giovanni Boero\textsuperscript{\textcolor{blue}{1},\textcolor{blue}{2}}\orcid{0000-0003-4913-3114}, 
Hernán Furci\textsuperscript{\textcolor{blue}{1}}\orcid{0000-0003-0595-4180}
}

\affil{\textsuperscript{1}Microsystems Laboratory, \'Ecole Polytechnique Fédérale de Lausanne (EPFL), 1015 Lausanne, Switzerland}

\affil{\textsuperscript{2}Center for Quantum Science and Engineering, \'Ecole Polytechnique Fédérale de Lausanne (EPFL), 1015 Lausanne, Switzerland}

\affil{\textsuperscript{*}Author to whom any correspondence should be addressed.}

\email{\textcolor{blue}{andre.chatel@epfl.ch}}


\begin{abstract}
\vspace{3mm} 
\justifying
\noindent
The rising complexity of cutting-edge cryogenic systems is currently imposing challenging technical constraints to the monitoring of ultra-cold temperatures through standard commercially available sensors. Among different alternative technologies, superconducting microwave resonators have been recently investigated as ideal candidates for performing on-chip cryogenic thermometry, in reason of their intrinsically low power dissipation, typically large temperature sensitivities and excellent sub-$\text{\SI{}{\mu\kelvin}}$ resolution below $\text{\SI{10}{\kelvin}}$. In such a framework, through this study we aim at demonstrating the possibility to enhance the temperature performance of superconducting microwave resonators by means of surface nanostructuring. More specifically, different arrays of nanotrenches are strategically patterned on the inductive line of a $\text{\SI{1.3}{\giga\hertz}}$ planar resonator, by partially etching a Nb\textsubscript{50}Ti\textsubscript{50} thin film, in order to tune the critical transition of the material and, therefore, increase the curvature of the $f_\text{res}(T)$ response. Although the presence of such thinned superconducting regions introduces larger microwave losses, a $\text{\SI{1.5}{\kelvin}}$ decrease of $T_\text{C}$ is recorded, which directly translates into an enhancement of the temperature sensitivity by a factor $\text{\SI{10}{}}$, with respect to a reference non-nanostructured sensor. In particular, a maximum value of $df_\text{res}/dT \simeq \text{\SI{62}{\mega\hertz/\kelvin}}$, at $\text{\SI{4.2}{\kelvin}}$, is achieved for the device showing the largest nanotrench width of about $\text{\SI{350}{\nano\metre}}$. These results demonstrate that such a surface nanostructuring approach can be exploited as an effective design strategy to tune the $f_\text{res}(T)$ response of superconducting microwave resonators and enhance their temperature sensitivity around a target operating condition. Although this technique is found to introduce additional RF losses into the superconducting device, thereby limiting the achievable temperature resolution to the few-$\text{\SI{}{\mu\kelvin/\hertz^{1/2}}}$ range (at $\text{\SI{1}{\hertz}}$ and $\text{\SI{4.2}{\kelvin}}$), we believe that similar approaches could be further investigated for the development of future cryogenic sensors. 
\end{abstract}

\begin{multicols*}{2}
\justifying

\section{Introduction}

By looking back at the scientific and technological advancements achieved all along the last decades, it is evident that cryogenic systems have been constantly playing a role of primary importance. Indeed, many different ground-breaking technologies, such as magnetic resonance imaging (MRI) \cite{Pittard:1993, Patel:2017, Mukhatov:2023, Jimeno:2023}, superconducting magnets \cite{Schmuser:1991, Lebrun:2017, Bottura:2022} and superconducting radio-frequency (SRF) cavities \cite{Padamsee:2017, Gurevich:2023, Petersen:2026} used in high-energy particle accelerators, quantum computing architectures \cite{Wallraff:2004, Goppl:2008, Scarlino:2019, De_Palma:2024, Awschalom:2025} and high-resolution radiation detectors for astrophysics \cite{Audley:1993, Day:2003, LeDuc:2010, O'Connor:2019, Dibert:2022}, require operating temperatures well below $\text{\SI{120}{\kelvin}}$. The accurate temperature control and monitoring is, therefore, fundamental for guaranteeing the correct functioning of the system and, consequently, enabling physical phenomena solely occurring in cryogenic conditions. Nowadays, a broad variety of well-established sensors is commercially available, with resistance temperature detectors (RTDs), in the form of CERNOX\textsuperscript{\textregistered} \cite{Yeager:2001, Courts:2003, Ekin:2006, Courts:2014} or RuO\textsubscript{2} \cite{Holmes:1992, Courts:2008, Myers:2021} thin-films, and Si/Ge diodes \cite{Ekin:2006, Courts:2016} representing the current state of the art for cryogenic thermometry. As a matter fact, such devices are characterised by a large $\text{\SI{1}{}-\SI{300}{\kelvin}}$ operating range, ease of electronic readout, low cross-sensitivity with magnetic fields and a temperature resolution typically as good as $\text{\SI{100}{\mu\kelvin}}$, making them ideal candidates to accurately sense low-temperatures in large-scale apparatuses \cite{Ekin:2006}. Nevertheless, the rising complexity of cryogenic systems imposes challenging design constraints to the monitoring of ultra-cold temperatures through the aforementioned devices, because of their relatively large footprint, need for individual electrical routings and significant power dissipation. 

Among different technologies, superconducting materials are currently investigated as potential alternatives to perform high-resolution cryogenic thermometry, especially when considering miniaturised quantum electro and thermodynamics circuits operating in the sub-kelvin regime. While recent researches in superconductivity have largely focused on the synthesis and characterization of novel superconducting materials \cite{Uebe:2024, Zhou:2025, Wang:2026}, significant efforts have also been dedicated to the engineering of superconducting microwave devices for thermometric applications. Indeed, different approaches have been recently proposed, ranging from tunnel \cite{Benoît:1993, Golubev:1999, Feshchenko:2015} and Josephson junctions \cite{Faivre:2014, Wang:2018, Zgirski:2018, Lvov:2025}, with a temperature resolution as low as $\text{\SI{2}{\mu\kelvin/\hertz^{1/2}}}$, around $\text{\SI{300}{\milli\kelvin}}$ \cite{Faivre:2014}, to CMOS-compatible and ultra-miniaturised superconducting thin-films integrated within the architecture of quantum processors \cite{Noah:2024}. Besides devices exclusively operating in DC, superconducting microwave resonators have also been proven to represent excellent candidates for multiplexing the temperature readout of several sensors, with quality factors larger than $10^{4}$ and sub-$\text{\SI{}{\mu\kelvin}}$ resolution, in parallel with the cryogenic operation of the device under test (DUT) \cite{Wheeler:2020, Chatel:2025}.

Following the results reported in a recent work by our group \cite{Chatel:2025}, where a high-performance cryogenic thermometer, characterised by sub-$\text{\SI{}{\mu\kelvin}}$ temperature resolution and based on a frequency-multiplexed readout of different superconducting microwave resonators, has been demonstrated, here we investigate the influence of surface nanostructuring on the temperature performance of similar sensors. More specifically, this study originates as an attempt to develop an additional design tool for tuning, and eventually enhancing, the temperature sensitivity of thin-film superconducting resonators. Indeed, these devices typically experience  a non-linear degradation of such a figure-of-merit (FOM) for temperatures receding from the critical transition $T_\text{C}$, in agreement with the temperature behaviour of their kinetic inductance $L_\text{K}(T)$ \cite{Watanabe:1994, Hein:2001, Frunzio:2005, Wandui:2020}, with this latter representing the main contribution to the electrodynamic response of thin-film superconducting resonators \cite{Clem:2012, Kubo:2022}. Our findings demonstrate that, by strategically nanostructuring the inductive line of a superconducting thin-film resonator, through an array of partially etched trenches, it is possible to tailor the critical temperature of the material and, therefore, accentuate the curvature of the $f_\text{res}(T)$ response at the specific operating temperature. Although such thinned superconducting regions act as an additional source of microwave losses (i.e. inducing an increase of the quasiparticles density) and, as a drawback, a degradation of the quality factor is consequently expected, a $\text{\SI{10}{}}$-fold enhancement of the temperature sensitivity is recorded, for instance at $\text{\SI{4.2}{\kelvin}}$, between a reference non-nanostructured resonator and the maximally nanostructured one.

\section{Methods}

From a general perspective, superconducting weak-links are typically associated to controlled and well-engineered material discontinuities, usually in the form of an insulating (SIS), normal-conducting (SNS) or weakened-superconducting (SS'S) layer inter-imposed between two superconducting banks \cite{Likharev:1979}, which can locally suppress the order parameter at the scale of few coherence lengths, but still ensuring a supercurrent to flow across the barrier. In the last decades, such structures have been widely exploited to realise advanced cryogenic devices based on the Josephson effect \cite{Josephson:1962,, Fagaly:2006, Bi:2024}, as, for instance, superconducting quantum interference devices (SQUIDs) \cite{Giazotto:2010, Qu:2012, Sharon:2016} and qubits \cite{Koch:2007, Schreier:2008, Barends:2014, Kjaergaard:2020}. More specifically, the possibility of Cooper pairs to travel inside a weaker-conducting material, as well as normal electrons to diffuse into the superconducting material and contribute to the electronic transport in the form of quasiparticles, is a well-known phenomenon commonly referred as proximity effect. Additionally, several studies have reported proximity effect to show a gradual disappearance when increasing the distance between the two superconducting contacts (i.e. the thickness of the weaker-conducting material in-between), with critical transition parameters, such as $T_\text{C}$ and $J_\text{C}$, shifting to lower values \cite{Meissner:1960, Kircher:1968, Clarke:1968}.

Considering, now, the application of similar features for cryogenic thermometry, if such nanostructures were strategically embedded in the high current density areas of a thin-film microwave superconducting resonator, the proximity effect should, therefore, generate a temperature-related effect on the frequency response of the device, additional to the intrinsic kinetic inductance one. In a first attempt to demonstrate this idea, our group has initially investigated the surface nanostructuring of a metal-superconductor bilayer (e.g. Pt-Nb\textsubscript{50}Ti\textsubscript{50} and Al-Nb\textsubscript{50}Ti\textsubscript{50}), where superconducting nano-islands have been bridged by means of the proximity effect induced by a normal metal film underneath. However, a degradation of the quality factor, and consequently a worsening of the noise equivalent temperature ($NET$), of more than two orders of magnitude, with respect to purely superconducting devices, has been observed. Such significantly higher microwave losses, probably due to the presence of interface defects and an excessive injection of quasiparticles, from the normal metal into the superconducting layer, have motivated us to investigate a simpler, cleaner and interface-free alternative approach, discussed all along this work, for which the weakened-superconducting regions are realised by partially etching the superconducting thin-film. As a matter of fact, several studies have reported the superconducting parameters of a thin-film to degrade with a reduction of the material thickness \cite{Takeda:1999, Ilin:2004, Gubin:2005, Inomata:2009, Bretz-Sullivan:2022, Zhu:2024}, which might still be exploited, in an way analogous to proximity effect, for locally suppressing the order parameter and, thus, introduce weakened-superconducting regions in the form of partially etched nanotrenches.

\begin{figure*}[t!]
\includegraphics[width=\linewidth]{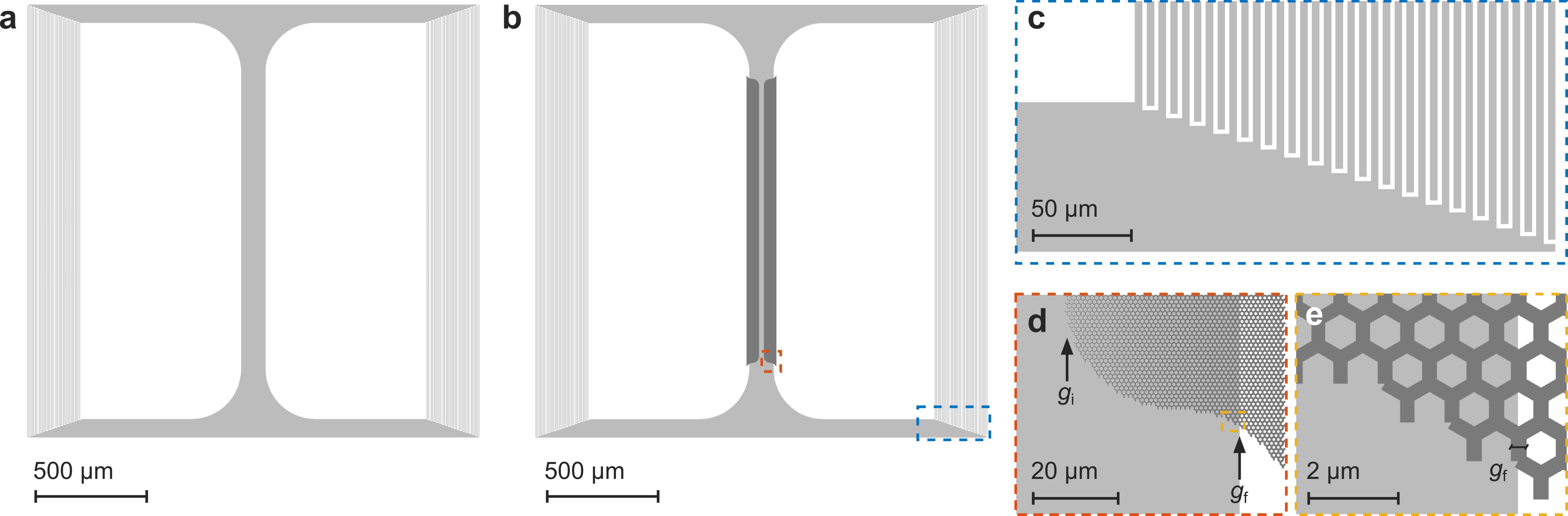}
\caption{\justifying Design schematic representing the layout of the nanostructured superconducting resonator. a) Reference resonator, with no surface nanostructuring along the main central inductive line and two lateral IDCs. b) Nanostructured resonator, with two arrays of weakened-superconducting regions laterally extending across the main central inductive line. The colors indicate two consecutive patterning layouts: in light grey, the fully-etched resonator pattern ($\text{\SI{100}{\nano\metre}}$ thick), while, in dark grey, the partially-etched array of thinner nanotrenches  ($\text{\SI{20}{\nano\metre}}$ thick). c) Zoom on the region of the IDC fingers. d) Zoom on the bottom extension of the inductive nanostructures array. The two arrows show the position of the minimum $g_\text{i}$-wide and maximum $g_\text{f}$-wide nanostructures across the main inductance line. e) Zoom on the hexagonal islands and interconnecting nanotrenches ($g_\text{f} = \text{\SI{350}{\nano\metre}}$, for this specific design). The arrays laterally exceed the inductive line by $\text{\SI{10}{\mu\metre}}$; this is designed on purpose, in order to avoid any possible misalignment that might occur between the two subsequent patternings. At the end of the processing of the fabricated devices, this area will correspond to the bare substrate.}
\label{MAIN_FIG_01}
\makeatletter
\let\save@currentlabel\@currentlabel
\edef\@currentlabel{\save@currentlabel(a)}\label{MAIN_FIG_01:a}
\edef\@currentlabel{\save@currentlabel(b)}\label{MAIN_FIG_01:b}
\edef\@currentlabel{\save@currentlabel(c)}\label{MAIN_FIG_01:c}
\edef\@currentlabel{\save@currentlabel(d)}\label{MAIN_FIG_01:d}
\edef\@currentlabel{\save@currentlabel(e)}\label{MAIN_FIG_01:e}
\edef\@currentlabel{\save@currentlabel(a-b)}\label{MAIN_FIG_01:a-b}
\edef\@currentlabel{\save@currentlabel(d-e)}\label{MAIN_FIG_01:d-e}
\makeatother
\end{figure*}

\subsection{Design of the nanostructured resonators}

Among several design options, lumped-elements microwave resonators offer the possibility to perform a fully wireless electromagnetic coupling with standard Cu grounded coplanar waveguides (GCPWs). In particular, a planar geometry similar to the one reported in \cite{Russo:2025} is chosen for the realisation of the nanostructured resonators, with an isolated central inductive line (length: $\text{\SI{1.4}{\milli\metre}}$, width: $\text{\SI{100}{\mu\metre}}$) laterally separated from two interdigitated capacitors (IDCs) by a $\text{\SI{650}{\mu\metre}}$ distance. Two typologies of resonators are designed, as depicted in \autoref{MAIN_FIG_01:a-b}: a first reference one, with no nanostructured arrays, and a set of three different resonators, maintaining the same overall geometry, but implementing different distributions of the nanofeatures. Moreover, the $\text{\SI{36}{}}$ fingers composing each IDC show an average length of $\text{\SI{1.67}{\milli\metre}}$, a $\text{\SI{4}{\mu\metre}}$ width and $\text{\SI{2}{\mu\metre}}$ gap size [see \autoref{MAIN_FIG_01:c}]. In first approximation, the geometric inductance of such a structure is calculated by recurring to an analytical formula \cite{Rosa:1908}
\begin{equation}
\label{EQUATION_MAIN_00a}
L_\text{G} \simeq \frac{\mu_0 l}{2\pi} \ln\biggl(\frac{2l}{w+t}\biggr)
\end{equation}
where $l$, $w$ and $t$ represent respectively the length, width and thickness of the rectangular cross-section of the line, resulting in an estimated value of about $L_\text{G} \simeq \text{\SI{1}{\nano\henry}}$. Moreover, the total value of the capacitance is also determined, through the analytical expression reported in \cite{Igreja:2004} and based on conformal-mapping techniques \cite{Simons:2001, Pozar:2012}, leading to an estimation of $ C_\text{G} \simeq \text{\SI{15}{\pico\farad}}$ for the $\text{\SI{72}{}}$ fingers composing the two IDCs. These two results can, then, be exploited to compute a first approximated resonance frequency as
\begin{equation}
\label{EQUATION_MAIN_00b}
f_\text{res} = \frac{1}{2\pi\sqrt{C_\text{G}L_\text{G}}} \simeq \text{\SI{1.3}{\giga\hertz}}
\end{equation}
Such calculations are further validated by means of a finite element method (FEM) simulation, implemented through COMSOL Multiphysics\textsuperscript{\textregistered} and meant to verify the correct anti-symmetric inductive coupling of such a geometry with a standard microwave coplanar waveguide (CPW) [see the \textcolor{blue}{supplementary data, S1}].

Considering the surface nanostructuring, two lateral arrays are designed across the main inductance line, as shown in \autoref{MAIN_FIG_01:d-e}. Several superconducting hexagonal islands are, therefore, generated, with a $\text{\SI{600}{\nano\metre}}$ side and weakened superconducting nanotrenches interconnecting them. The width extension of such nanotrenches linearly decreases, from a given maximum $g_\text{f}$ located at the lateral side of the inductive line, down to a minimum $g_\text{i}$ of $\text{\SI{50}{\nano\metre}}$, $\text{\SI{40}{\mu\metre}}$ far from the outermost edge. Such a gradual transition on the value of the nanotrenches is implemented in order to provide a smooth microwave current distribution across the the central inductive line. Moreover, the large extension of the main line promotes crowding of the microwave current towards its edges \cite{Simons:2001}. As a consequence, the wider and more laterally positioned nanotrenches are expected to act more actively, providing the dominant contribution to the observed temperature-induced resonance frequency shift. In order to study such a possible effect, three different resonators are also designed, in addition to the reference device, with three different nanotrenches distributions, showing, at the edges, $g_\text{f}$ values of $\text{\SI{150}{\nano\metre}}$, $\text{\SI{250}{\nano\metre}}$ and $\text{\SI{350}{\nano\metre}}$ respectively.

\subsection{Nanostructuring Nb\textsubscript{50}Ti\textsubscript{50} resonators on sapphire}

The nanostructured resonators are patterned from a $\text{\SI{100}{\nano\metre}}$ thick Nb\textsubscript{50}Ti\textsubscript{50} film, DC-sputtered on a $\text{\SI{650}{\mu\metre}}$ thick sapphire wafer. More specifically, the process flow consists of two patterning sequences, both based on SF\textsubscript{6}/CHF\textsubscript{3} plasma etching, with the first one relying on electron-beam lithography (EBL) to pattern the array of nanotrenches, while the second one exploits direct laser writing (DLW) for defining the planar geometry of the resonator all around the nanostructured area. The technical details concerning the micro and nano-fabrication of such devices are fully described in the \textcolor{blue}{supplementary data, S2}. Dealing with the reasons behind the choice of the materials, sapphire is chose as an ideal substrate for its thermal \cite{Dobrovinskaya:2009, Brown:2010} and electronic \cite{Buckley:1994, Krupka:1999, Pogue:2012} properties in cryogenic conditions. Indeed, this material is commonly exploited in low-temperature microwave applications as it shows a thermal conductivity  $\kappa_\text{T}$ as high as $\text{\SI{1000}{\watt/(\metre\cdot\kelvin)}}$, a large dielectric constant $\epsilon_\text{r} \simeq \text{\SI{10}{}}$ and an almost negligible loss factor  $\epsilon'' < 10^{-9}$. Moreover, the superconducting thin-film resonators are made of Nb\textsubscript{50}Ti\textsubscript{50} in reason of the compatibility between its bulk critical temperature $T_\text{C} \simeq \text{\SI{10}{\kelvin}}$ \cite{Benvenuti:1991, Charifoulline:2006, Zhang:2019} and our cryogenic setup \cite{Russo:2023}, as well as a large critical field of about $\text{\SI{15}{\tesla}}$ \cite{Bychkov:1981, Bottura:2000, Ghigo:2023}, useful to mitigate the cross-sensitivity of the devices to $B$-fields.

\begin{figure}[H]
\includegraphics[width=\linewidth]{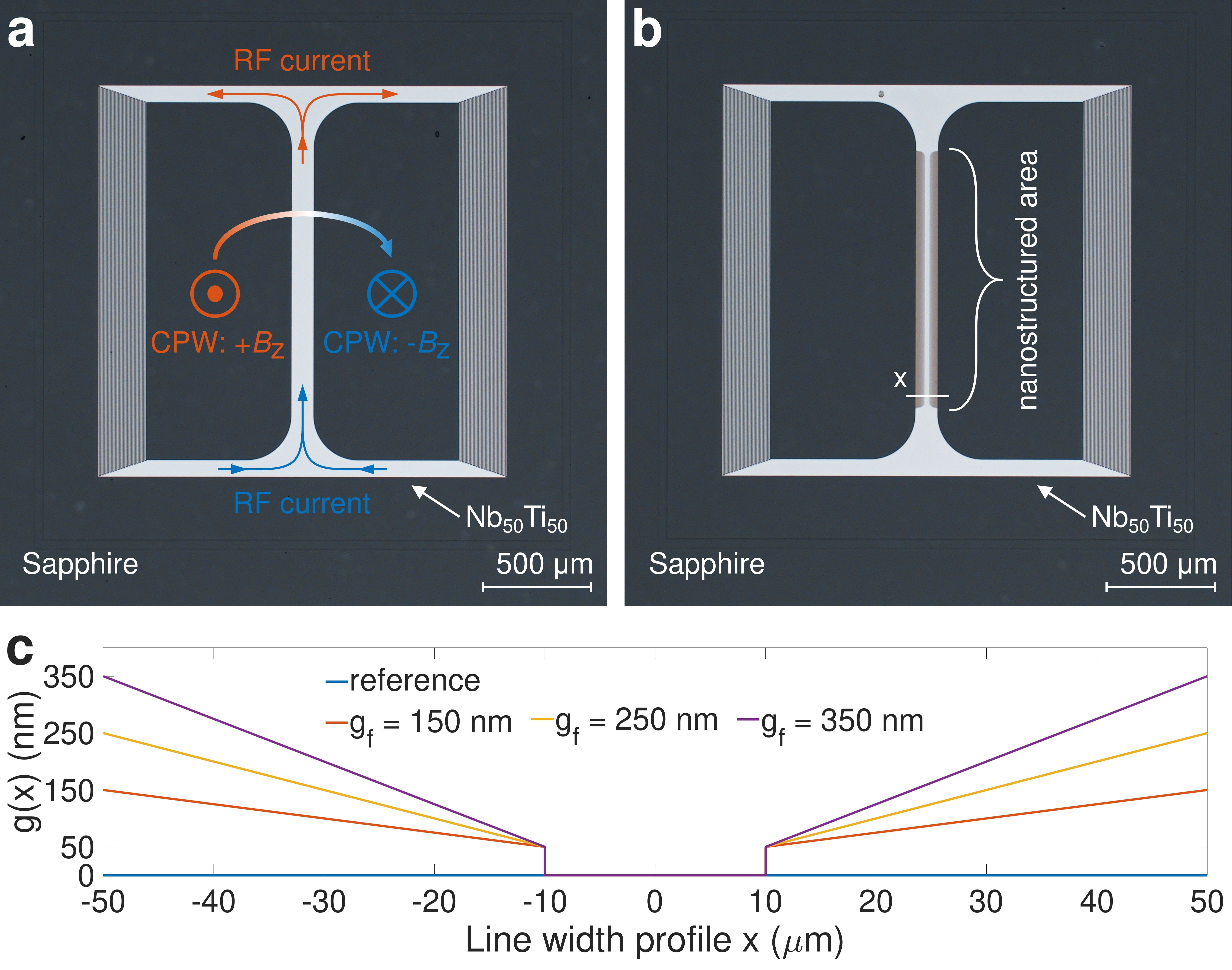}
\caption{\justifying Optical microscope pictures of the fabricated nanostructured resonators. a) Reference resonator. The coloured arrow indicates the direction of the coupling antisymmetric $B$-field generated by an excitation CPW. b) $g_\text{f} = \text{\SI{350}{\nano\metre}}$ nanostructured resonator, showing the arrays of nanotrenches in correspondence of the central inductive line. c) Distribution of the nanotrenches width values across the central line, for all the four DUTs.}
\label{MAIN_FIG_02}
\makeatletter
\let\save@currentlabel\@currentlabel
\edef\@currentlabel{\save@currentlabel(a)}\label{MAIN_FIG_02:a}
\edef\@currentlabel{\save@currentlabel(b)}\label{MAIN_FIG_02:b}
\edef\@currentlabel{\save@currentlabel(c)}\label{MAIN_FIG_02:c}
\edef\@currentlabel{\save@currentlabel(a-b)}\label{MAIN_FIG_02:a-b}
\makeatother
\end{figure}

\begin{figure}[H]
\includegraphics[width=\linewidth]{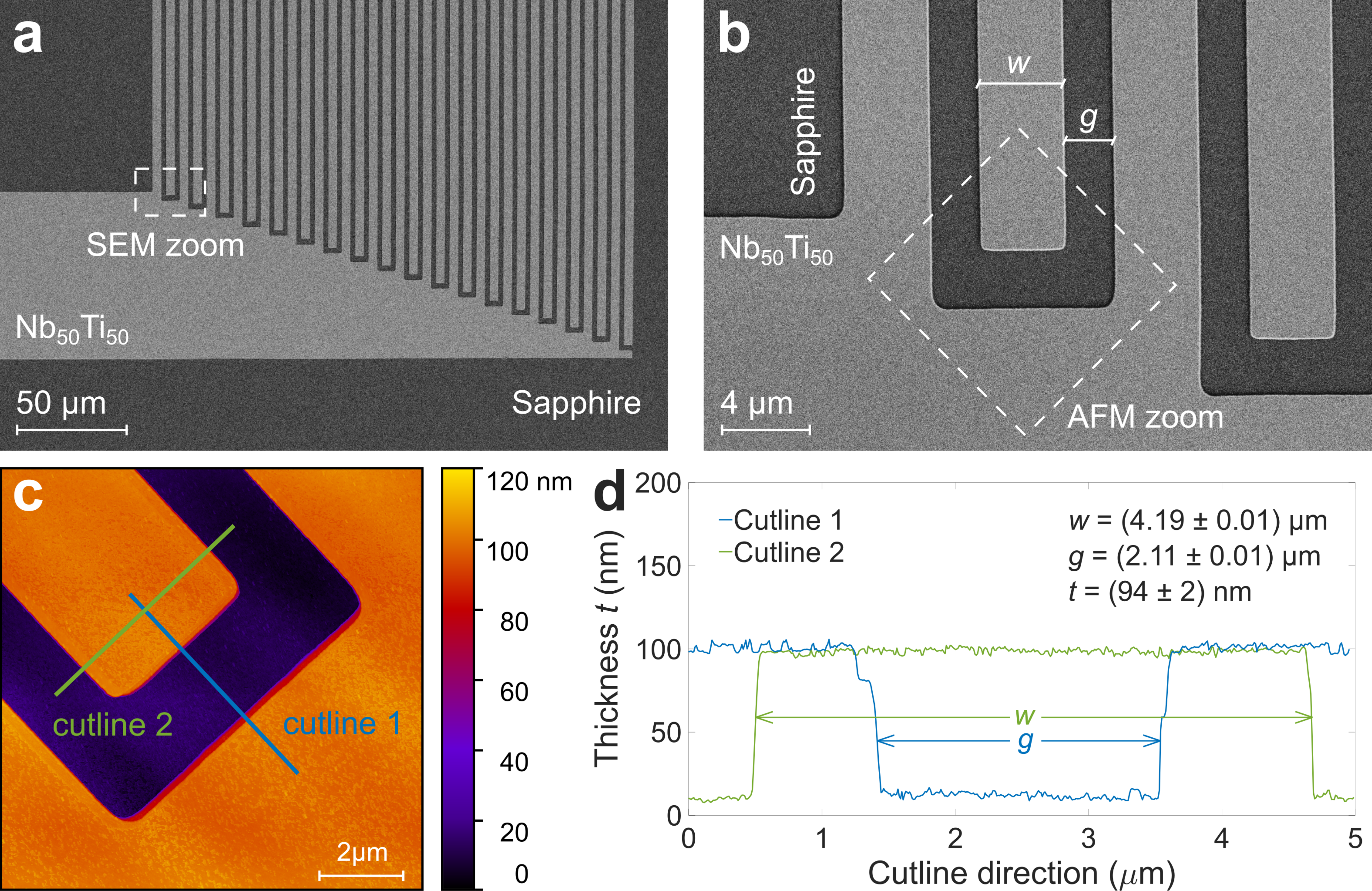}
\caption{\justifying Characterisation of the resonator patterning. a) SEM picture highlighting the bottom-left corner of the IDC. b) Zoom on the details of the IDC fingers (width: $\text{\SI{4}{\mu\metre}}$, gap: $\text{\SI{2}{\mu\metre}}$). c) AFM scan of a single finger. The thickness of the DC-sputtered Nb\textsubscript{50}Ti\textsubscript{50} film is supposed to be $\text{\SI{100}{\nano\metre}}$, by design. d) AFM cutlines along width and gap of the previous finger.}
\label{MAIN_FIG_03}
\makeatletter
\let\save@currentlabel\@currentlabel
\edef\@currentlabel{\save@currentlabel(a)}\label{MAIN_FIG_03:a}
\edef\@currentlabel{\save@currentlabel(b)}\label{MAIN_FIG_03:b}
\edef\@currentlabel{\save@currentlabel(c)}\label{MAIN_FIG_03:c}
\edef\@currentlabel{\save@currentlabel(d)}\label{MAIN_FIG_03:d}
\edef\@currentlabel{\save@currentlabel(a-b)}\label{MAIN_FIG_03:a-b}
\edef\@currentlabel{\save@currentlabel(c-d)}\label{MAIN_FIG_03:c-d}
\makeatother
\end{figure}

\begin{figure}[H]
\includegraphics[width=\linewidth]{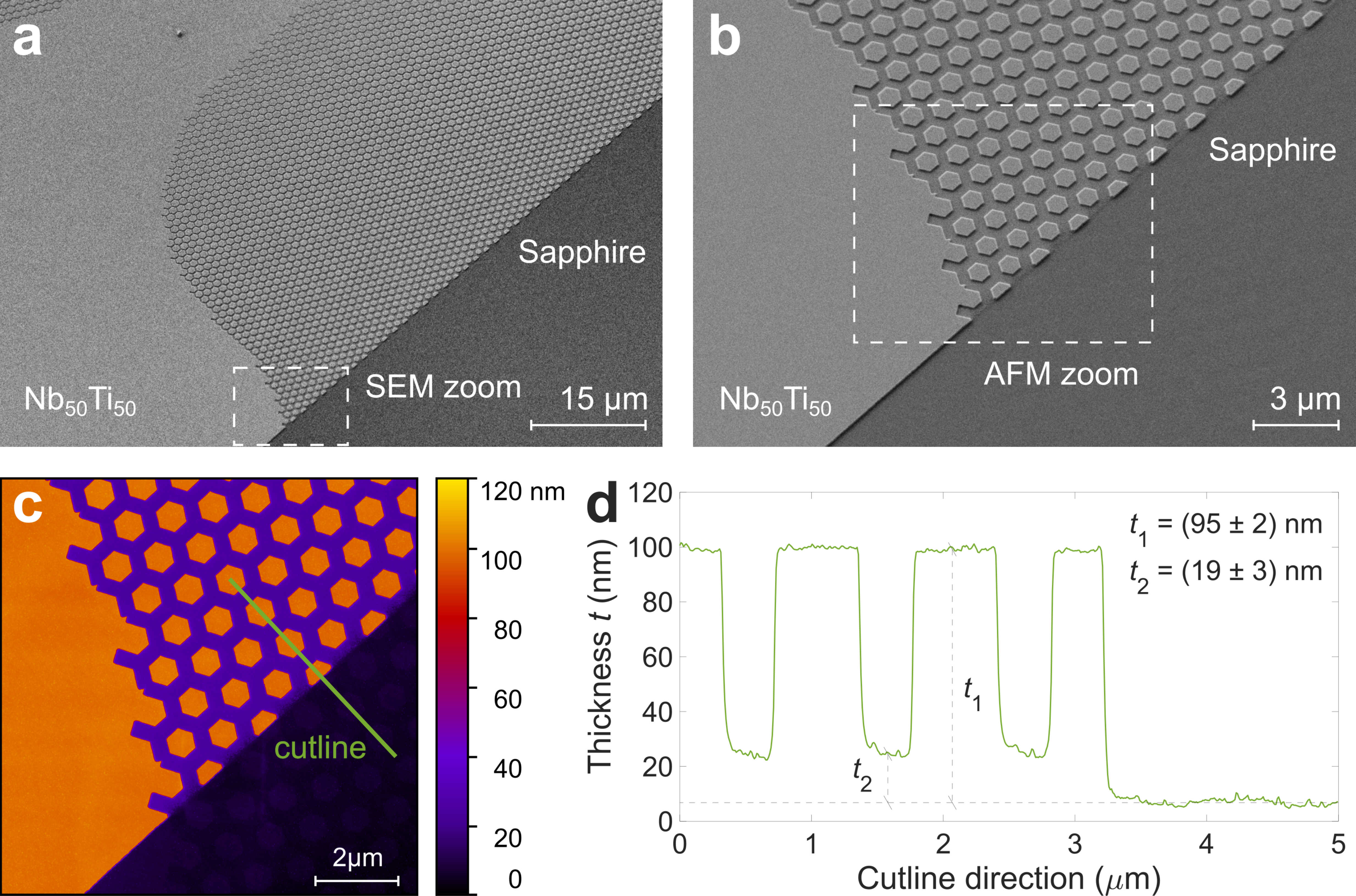}
\caption{\justifying Characterisation of the partially etched Nb\textsubscript{50}Ti\textsubscript{50} regions. a) SEM picture highlighting one of the corners of the nanotrenches array located on the main inductive line of the resonator. b) Zoom on the details of the largest outermost trenches (in this case, $g_\text{f} = \text{\SI{350}{\nano\metre}}$). c) AFM scan of the outermost trenches. The $t_\text{2}$ thickness of the remaining Nb\textsubscript{50}Ti\textsubscript{50} film, after a partial plasma etching, is supposed to be between $\text{\SI{20}{\nano\metre}}$ to $\text{\SI{30}{\nano\metre}}$, by process design. d) AFM cross-sectional cutline along several thinned Nb\textsubscript{50}Ti\textsubscript{50} trenches.}
\label{MAIN_FIG_04}
\makeatletter
\let\save@currentlabel\@currentlabel
\edef\@currentlabel{\save@currentlabel(a)}\label{MAIN_FIG_04:a}
\edef\@currentlabel{\save@currentlabel(b)}\label{MAIN_FIG_04:b}
\edef\@currentlabel{\save@currentlabel(c)}\label{MAIN_FIG_04:c}
\edef\@currentlabel{\save@currentlabel(d)}\label{MAIN_FIG_04:d}
\edef\@currentlabel{\save@currentlabel(a-b)}\label{MAIN_FIG_04:a-b}
\edef\@currentlabel{\save@currentlabel(c-d)}\label{MAIN_FIG_04:c-d}
\makeatother
\end{figure}

Once the resonators have been patterned and diced into $\text{\SI{10}{\milli\metre}}\times\text{\SI{10}{\milli\metre}}$ squared chips, the fabrication quality is assessed by optical microscopy, scanning electron microscopy (SEM) and atomic force microscopy (AFM). In \autoref{MAIN_FIG_02:a-b}, we report optical pictures of the sensors, including the reference resonator without any surface nanostructuring, as well as the $g_\text{f} = \text{\SI{350}{\nano\metre}}$ one. In particular, as previously mentioned,\hfill the\hfill distribution\hfill of\hfill the\hfill partially\hfill etched\hfill Nb\textsubscript{50}Ti\textsubscript{50}\hfill re-
\noindent gions is designed in order to linearly increase their trench width from a minimum $g_\text{i} = \text{\SI{50}{\nano\metre}}$, $\text{\SI{10}{\mu\metre}}$ far from the centre of the main inductance line, up to a maximum value $g_\text{f}$, of $\text{\SI{150}{\nano\metre}}$, $\text{\SI{250}{\nano\metre}}$ and $\text{\SI{350}{\nano\metre}}$, in correspondence of the outermost lateral edges [see \autoref{MAIN_FIG_02:c}]. Furthermore, from the point of view of the planar resonator, the processing critical dimensions are the ones related to the IDC fingers, as shown in \autoref{MAIN_FIG_03:a-b}, whose width and gap sizes should be, respectively, equal to $\text{\SI{4}{\mu\metre}}$ and $\text{\SI{2}{\mu\metre}}$. By means of AFM scans, it is possible to quantify the geometrical deviations from the design, as well as determine the exact thickness of the thin-film device. As described by \autoref{MAIN_FIG_03:c-d}, a design transfer inaccuracy of less than $\text{\SI{200}{\nano\metre}}$ is estimated, consistently with our previous patterning of Nb\textsubscript{50}Ti\textsubscript{50} thin-films \cite{Chatel:2025}, while the resulting thickness of the sputtered material turns out to be $\text{\SI{94\pm2}{\nano\metre}}$. The same characterisation procedure can be applied as well to inspect the patterning of the nanotrenches on the resonator main inductive line [see \autoref{MAIN_FIG_04:a-b}]. AFM scans are, therefore carried out with the ultimate goal of describing the morphology of the nanostructured array of partially etched trenches. As it is possible to notice from \autoref{MAIN_FIG_04:c-d}, the outermost trenches turn out to be properly defined by the plasma etching process, while the thickness of such features can be estimated to be in the order of $\text{\SI{19\pm3}{\nano\metre}}$.

\subsection{Preliminary tests in liquid helium}

Before further proceeding with the description of the experimental results, it is important to mention that the resonators are also preliminarily characterised in liquid He (LHe) at $\text{\SI{4.2}{\kelvin}}$, in order to first determine optimal excitation conditions, in terms microwave power and coupling distance. In such a perspective, a Cu GCPW, similar to the one we previously designed in \cite{Chatel:2025}, is used to excite the devices through the\hfill microwave\hfill power\hfill delivered\hfill by\hfill a\hfill vector\hfill network\hfill analyser
\noindent (VNA) connected in transmission to the cryogenic printed circuit board (PCB). The results concerning such an analysis are described in detail in the \textcolor{blue}{supplementary data, S3}. In particular, an excitation power of $\text{\SI{-40}{\dBm}}$ is selected for further cryogenic characterisations, as it avoids the generation of non-linear effects and self-heating phenomena. Additionally, the coupling distance between the resonating chip and the CPW is varied by 3D-printing different polyactide (PLA) spacers. Among the investigated configurations, we select a coupling distance of $\text{\SI{0.75}{\milli\metre}}$, as it corresponds to a slightly under-coupled conditions, which provides loaded quality factor values typically larger than $\text{\SI{5e3}{}}$ and a high signal-to-noise ratio (SNR).

\section{Results and discussion}

In the following, we describe the different cryogenic experiments carried out to study the effects of surface nanostructuring on the temperature performance of the Nb\textsubscript{50}Ti\textsubscript{50} superconducting thin-film resonators. First, a preliminary DC characterisation is performed on several 4-wire test structures, each comprising of a single partially etched nanotrench, with the aim of investigating the influence of its width on the superconducting critical transition. In a second time, the $f_\text{res}(T)$ calibration curves are obtained for the different fabricated resonators, allowing to extract the sensitivity values across the operating temperature range. Finally, the estimation of the temperature resolution is performed in liquid He at $\text{\SI{4.2}{\kelvin}}$, by exploiting a low-noise microwave instrumentation setup relying on frequency modulation (FM) techniques.

\subsection{4-wire DC characterisation of single-nanotrenches}

\begin{figure}[H]
\includegraphics[width=\linewidth]{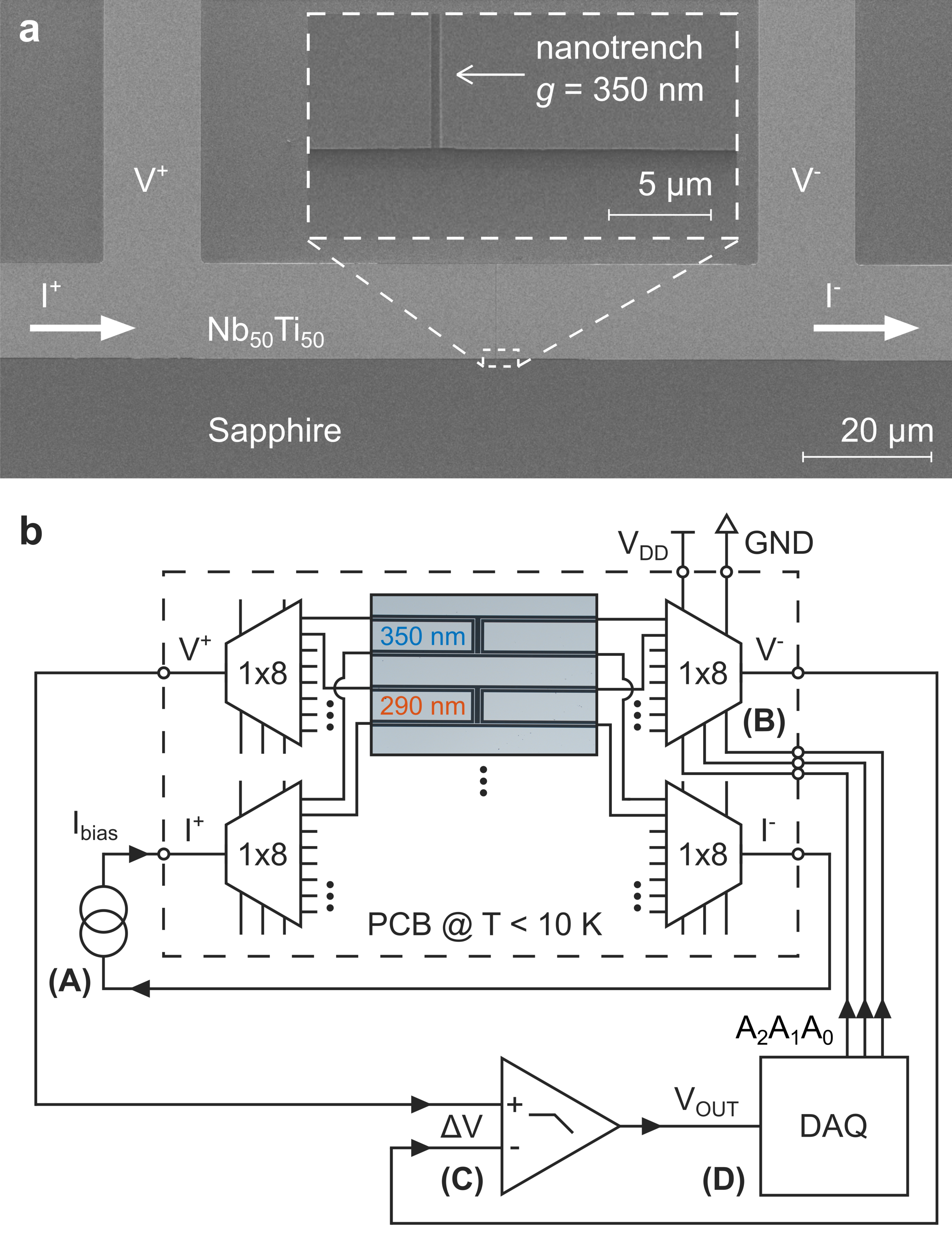}
\caption{\justifying 4-wire DC characterisation of the single-nanotrench structures. a) SEM picture showing the geometry of a 4-wire DC test structure, with a zoom on the single-nanotrench line (in this case, the $\text{\SI{350}{\nano\metre}}$ wide feature). b) Instrumentation setup exploiting a $1\times8$ cryogenic multiplexing routing to perform the 4-wire DC characterisation of single-nanotrench lines with variable width.}
\label{MAIN_FIG_05}
\makeatletter
\let\save@currentlabel\@currentlabel
\edef\@currentlabel{\save@currentlabel(a)}\label{MAIN_FIG_05:a}
\edef\@currentlabel{\save@currentlabel(b)}\label{MAIN_FIG_05:b}
\makeatother
\end{figure}

In order to study the effects of the nanotrench width on the superconducting critical transition of such partially etched Nb\textsubscript{50}Ti\textsubscript{50} regions, we simultaneously pattern a chip containing eight different 4-wire DC test structures on the same sapphire wafer from which the resonators are fabricated. In particular, each 4-wire structure includes a single partially etched region extending all across its whole width. Different nanotrenches $\text{\SI{50}{\nano\metre}}$, $\text{\SI{110}{\nano\metre}}$, $\text{\SI{170}{\nano\metre}}$, $\text{\SI{230}{\nano\metre}}$, $\text{\SI{290}{\nano\metre}}$ and $\text{\SI{350}{\nano\metre}}$ wide are, therefore, patterned and studied, together with a reference device consisting of a clean, non-nanostructured, 4-wire structure. An extra copy of the tiniest $\text{\SI{50}{\nano\metre}}$ nanotrench is added to the set, mainly exploited during the microfabrication phase to simply check the electrical connection between the two superconducting banks, after the partial etch of the nanostructure.  More specifically, the geometry of such a planar device consists of a $\text{\SI{100}{\mu\metre}}$ long resistive wire, of $\text{\SI{15}{\mu\metre}} \times \text{\SI{100}{\nano\metre}}$ rectangular cross-section, where a single nanotrench, extending all the across its line-width, is patterned at the centre, equally distanced from each lateral voltage reading line [see \autoref{MAIN_FIG_05:a}]. In addition to a visual inspection, AFM scans are also carried out for extracting the morphology of each single-nanotrench. The results of such an analysis are described in \textcolor{blue}{supplementary data, S4}, reporting a width deviation, from the design target values, always lower than $\text{\SI{10}{\nano\metre}}$, while a residual Nb\textsubscript{50}Ti\textsubscript{50} film thickness of about $\text{\SI{20}{\nano\metre}}$ is measured, consistent with the morphological characterisation of the nanostructured resonators.

The cryogenic DC characterisation of such 4-wire structures is performed inside the variable temperature insert (VTI) of a cryomagnet (from \textit{Cryogenic Ltd}), in zero-field conditions and for temperatures below $\text{\SI{10}{\kelvin}}$. In order to mitigate the cross-sensitivity of the resonators to external magnetic fields, a CRYOPHY\textsuperscript{\textregistered} shield is used to encapsulate the PCB at the cryogenic stage. Moreover, two couples of CERNOX\textsuperscript{\textregistered} RTDs and resistive heaters are exploited to perform the temperature control and monitoring, with the VTI temperature typically set $\text{\SI{200}{\milli\kelvin}}$ below the value targeted for the DUT \cite{Chatel:2025}. In \autoref{MAIN_FIG_05:b}, we schematically depict the electronic setup exploited to perform such a characterisation. More in detail, a source measure unit (SMU) (\textbf{A}. \textit{Keithley 2400})\hfill is\hfill set\hfill to\hfill bias\hfill the\hfill current\hfill branches\hfill of\hfill the\hfill different\hfill 4-
\noindent wire structures. The selection of each device is performed by programming a data acquisition (DAQ) board to control four different $1\times8$ CMOS bi-directional multiplexers (\textbf{B}. \textit{TMUX1108}), directly soldered on the PCB. The integration of such components at the cryogenic stage is implemented with the specific intention of increasing the experimental throughput at each cool-down. As a matter of fact, $\text{\SI{12}{}}$ DC pins are available from the probe of our cryostat, eventually limiting the characterisation to only three 4-wire structures per cool-down. In particular, such multiplexers, biased with a $\text{\SI{5}{\volt}}$ single power supply, are chosen for the very low quiescent current of about $\text{\SI{8}{\nano\ampere}}$, resulting in a total power consumption of $\text{\SI{4}{}} \times \text{\SI{40}{\nano\watt}}$. The differential voltage reading, corresponding to the selected 4-wire structure, is, then, amplified and low-pass filtered, with a $\text{\SI{30}{\hertz}}$ bandwidth (\textbf{C}. \textit{EG\&G Princeton Applied Research Model 5113}). The output signal is, finally, measured through a DAQ board (\textbf{D}. \textit{PCI-6052E}), whose acquisition time is set to $\text{\SI{100}{\milli\second}}$. Additionally, the extraction of the DC transition parameters, such as critical current density $J_\text{C}(T)$ and temperature $T_\text{C}(J)$, is performed by recurring to a fixed voltage threshold of about $\SI{1}{\mu\volt}$, corresponding to the RMS voltage noise level of the characterisation setup. More specifically, considering the $\SI{100}{\mu\metre}$ separation between the two voltage probes, this threshold\hfill determines\hfill an\hfill average\hfill electric\hfill field
\begin{figure}[H]
\includegraphics[width=\linewidth]{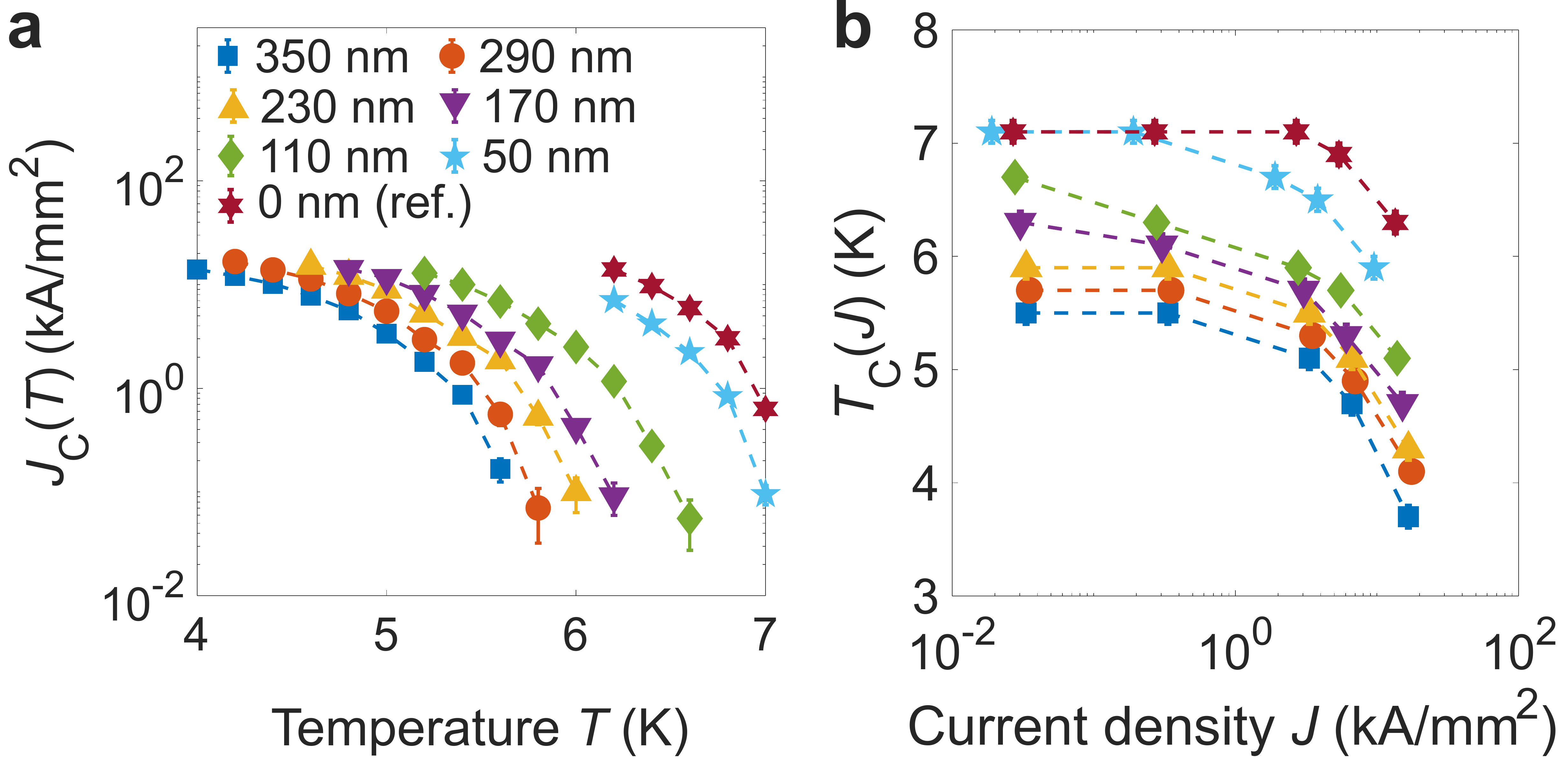}
\caption{\justifying $J_\text{C}$-$T_\text{C}$ superconducting critical transition for the single partially etched Nb\textsubscript{50}Ti\textsubscript{50} regions. a) Estimation of the critical current density $J_\text{C}$, for different nanotrench widths, performed through a current sweep at different operating temperatures. b) Estimation of the critical temperature $T_\text{C}$, for different nanotrench widths, performed through a temperature sweep at different biasing currents: the same colour legend from the previous $J_\text{C}$ vs $T$ study is also exploited in this plot.}
\label{MAIN_FIG_06}
\makeatletter
\let\save@currentlabel\@currentlabel
\edef\@currentlabel{\save@currentlabel(a)}\label{MAIN_FIG_06:a}
\edef\@currentlabel{\save@currentlabel(b)}\label{MAIN_FIG_06:b}
\makeatother
\end{figure}
\noindent criterion expressed as $E_\text{th}=\SI{100}{\mu\volt/\centi\metre}$.

The results concerning the overall $J_\text{C}$-$T_\text{C}$ transition, for the different single-nanotrenches, are reported in \autoref{MAIN_FIG_06}. When performing such a characterisation, with variable current and temperature, two distinguishable transitions are always recorded, with the first one being associated to the partially etched Nb\textsubscript{50}Ti\textsubscript{50} region, while the second one represents the collapsing of the overall superconducting film, as described in the \textcolor{blue}{supplementary data, S5}. In particular, it is possible to notice how an increasing nanotrench width has the effect of deteriorating the superconducting critical transition of the thinned Nb\textsubscript{50}Ti\textsubscript{50} trench, both in critical current density $J_\text{C}$ and temperature $T_\text{C}$. In such a sense, this behaviour is analogous to the one associated to more standard superconductor-normal-superconductor discontinuities, governed by proximity effect \cite{Meissner:1960, Kircher:1968, Clarke:1968}. In particular, an overall $T_\text{C}$ swing of about $\text{\SI{1.5}{\kelvin}}$, from the $\text{\SI{7.0}{\kelvin}}$ reference condition, is recorded from the evolution of such a parameter with respect to the increasing values of the nanotrench width. This result indicates that the gradual, and local, suppression of the superconducting order parameter, through such partially etched regions, can be actually engineered to realise temperature-sensitive devices.

\subsection{VTI determination of the temperature sensitivity}

The same experimental apparatus, previously described and relying on the VTI of our cryomagnet, is now exploited to perform the microwave characterisation of different nanostructured resonators. In particular, each device is excited, $\text{\SI{0.75}{\milli\metre}}$ far from the Cu GCPW, through a $\text{\SI{-40}{\dBm}}$ microwave power delivered by a vector network analyser (\textit{PicoVNA 108}). We record the complex $S_\text{21}$ transmission parameters, for all the fabricated resonators, while sweeping the temperature from $\text{\SI{3}{}}$ to $\text{\SI{7}{\kelvin}}$, with steps of $\text{\SI{200}{\milli\kelvin}}$. More in detail, the resonance frequency values at different temperatures, as well as the ones of the different quality factors, are derived by means of a complex plane circular fit to a notch-type\hfill transmission\hfill model,\hfill described\hfill by\hfill the\hfill well-known\hfill litera-
\begin{figure}[H]
\includegraphics[width=\linewidth]{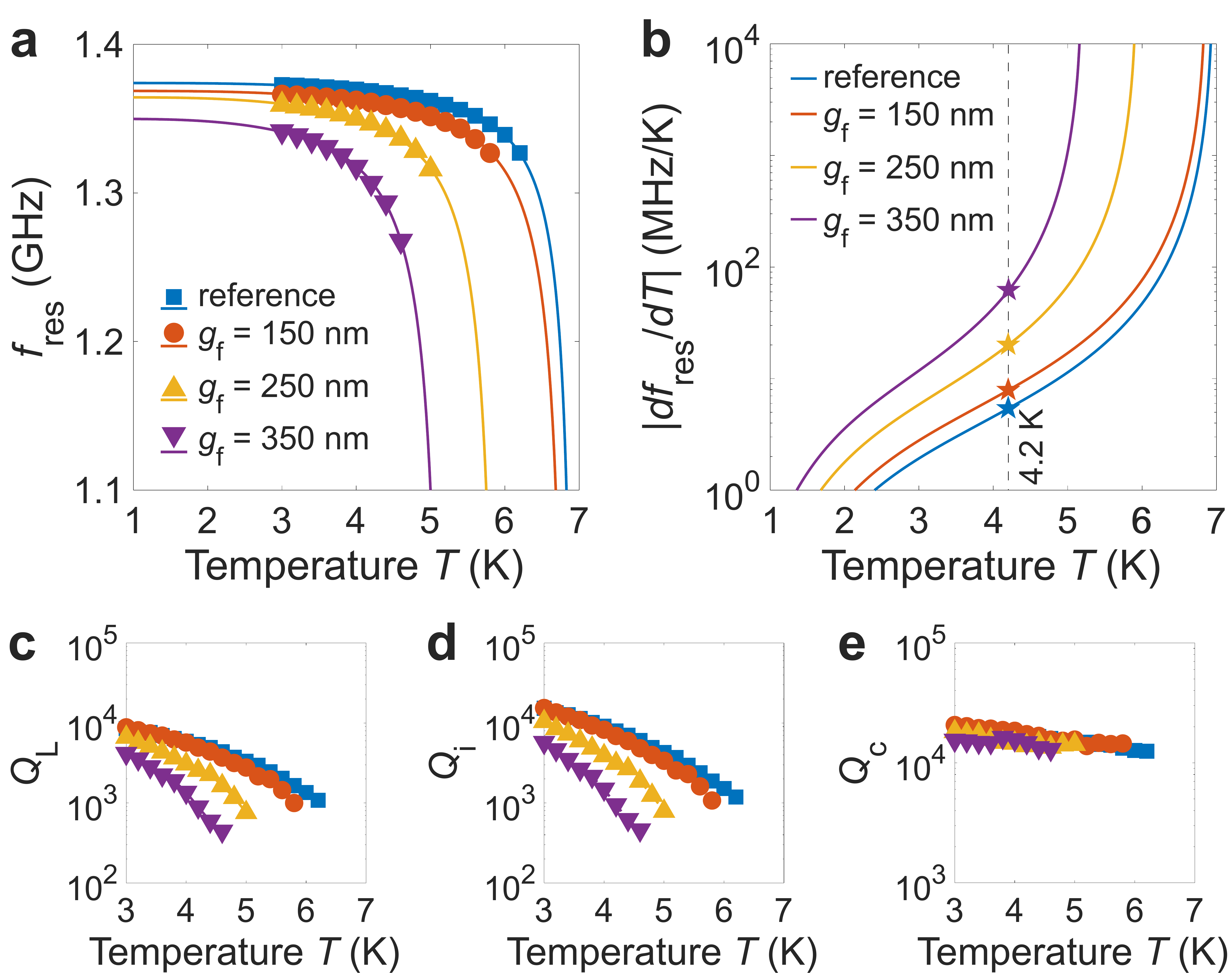}
\caption{\justifying Temperature response of the nanostructured resonators, for different nanotrench distributions. a) Resonance frequency shifts for the different nanotrench distributions: the $L_\text{K}(T)$-based model (solid lines) is exploited to fit the measured data points. b) Evolution of the temperature sensitivity, calculated as $df_{\text{res}}(T)/dT$ from the $L_\text{K}(T)$-based fitting model. The different pentagram points refer to the values estimated at $\text{\SI{4.2}{\kelvin}}$. c-e) Evolution of the loaded $Q_\text{L}$, internal $Q_\text{i}$ and coupled $Q_\text{c}$ quality factors: the colour legend is the same as for the previous graphs.}
\label{MAIN_FIG_07}
\makeatletter
\let\save@currentlabel\@currentlabel
\edef\@currentlabel{\save@currentlabel(a)}\label{MAIN_FIG_07:a}
\edef\@currentlabel{\save@currentlabel(b)}\label{MAIN_FIG_07:b}
\edef\@currentlabel{\save@currentlabel(c)}\label{MAIN_FIG_07:c}
\edef\@currentlabel{\save@currentlabel(d)}\label{MAIN_FIG_07:d}
\edef\@currentlabel{\save@currentlabel(e)}\label{MAIN_FIG_07:e}
\edef\@currentlabel{\save@currentlabel(c-e)}\label{MAIN_FIG_07:c-e}
\makeatother
\end{figure}
\noindent ture expression \cite{Gao:2008, Khalil:2012, Probst:2015}
\begin{equation}
\label{EQUATION_MAIN_01}
S_{\text{21}}(f) = A(f) \biggl( 1-\frac{(Q_\text{L}/|Q_\text{c}|)e^{i\phi}}{1+2iQ_\text{L}(f/f_{\text{res}}-1)} \biggr)
\end{equation}
where the term 
\begin{equation}
\label{EQUATION_MAIN_02}
A(f) = a e^{i(\alpha-2\pi\tau f)}
\end{equation}
is an environmental pre-factor, necessary to properly take into consideration the effect of the $a$ amplitude attenuation, $\alpha$ phase shift and $\tau$ time delay of the overall transmission line. Additionally, $f_{\text{res}}$ identifies the resonance frequency of the device, while eventual impedance mismatches are modelled by the factor $\phi$. Finally, the finite microwave losses affecting the resonator are modelled through the loaded $Q_\text{L}$ and coupling $Q_\text{c}$ quality factors. The internal Q-factor $Q_\text{i}$ can also be estimated by recurring to the constitutive relation $Q_\text{L}^{-1} = Q_\text{i}^{-1} + Q_\text{c}^{-1}$ \cite{Pozar:2012}.

\begin{table*}[t!]
\begin{threeparttable}
\caption{\justifying Critical temperature, geometric inductance, capacitance temperature sensitivity and internal Q-factor at $\text{\SI{4.2}{\kelvin}}$, for nanostructured resonators with different nanotrench distributions.}
\label{TABLE_MAIN_01}
\renewcommand{\arraystretch}{1.25}
\begin{tabular*}{\linewidth}{@{\hspace{3pt}\extracolsep{\fill}} cccccc @{\hspace{3pt}}}
\toprule
$g_\text{f}$ & $T_\text{C}$\tnote{a} & $L_\text{G}$\tnote{a} & $C_\text{G}$\tnote{a} & $df_\text{res}/dT$\tnote{b} & $Q_\text{i}$\tnote{c}\\
$\text{(nm)}$ & $\text{(\SI{}{\kelvin})}$ & $\text{(\SI{}{\nano\henry})}$ & $\text{(\SI{}{\pico\farad})}$ & $\text{(MHz/K)}$ & $\bigl(\text{$\times$10\textsuperscript{3}}\bigr)$\\
\midrule
$0$ (ref.) & $6.94\pm0.07$ & $0.99\pm0.08$ & $13\pm1$ & $5.42\pm0.03$ & $7.96$\\
$150$ & $6.84\pm0.08$ & $0.72\pm0.07$ & $18\pm2$ & $7.91\pm0.08$ & $6.88$\\
$250$ & $5.91\pm0.07$ & $0.69\pm0.06$ & $19\pm3$ & $20.1\pm0.3$ & $3.13$\\
$350$ & $5.18\pm0.06$ & $0.67\pm0.08$ & $20\pm3$ & $62\pm1$ & $0.95$\\
\bottomrule
\end{tabular*}
\begin{tablenotes}
\item[a] \footnotesize{Parameters extracted through the $L_\text{K}(T)$-based fitting model from \autoref{EQUATION_MAIN_03}.}
\item[b] \footnotesize{Determined as the temperature derivative of the fitting model from \autoref{EQUATION_MAIN_03}, evaluated at $\text{\SI{4.2}{\kelvin}}$.}
\item[c] \footnotesize{Extracted through the fit of the complex resonance dip, at $\text{\SI{4.2}{\kelvin}}$, to a circular notch-type model.}
\end{tablenotes}
\end{threeparttable}
\end{table*}

The temperature evolution of each resonance frequency is, therefore, plotted in \autoref{MAIN_FIG_07:a}, where it is possible to notice the larger curvature slope associated to the different nanotrench distributions. In order to calibrate the sensors against the temperature readings provided by the CERNOX\textsuperscript{\textregistered} thermistors, the resonance frequencies responses are additionally fitted to a kinetic inductance analytical expression \cite{Poole:1995, Kopnin:2001, Tinkham:2004, Annunziata:2010, Kubo:2020, Scarano:2024}, valid in the low frequency ($hf \ll \Delta_0$), thin-film ($t \ll \lambda_0$), low current density ($J \ll J_d$) and dirty limit ($l \ll \xi_0$) regimes. In particular, such a model can be expressed as 
\begin{subequations}
\label{EQUATION_MAIN_03}
\begin{equation}
f_{\text{res}}(T) = \frac{1}{2\pi\sqrt{C_\text{G}(L_\text{G}+L_\text{K}(T))}}
\label{EQUATION_MAIN_03a}
\end{equation}
\begin{equation}
L_\text{K}(T) = \frac{h R_{\square}(T_\text{C})}{2\pi^2 \Delta(T)} \frac{1}{\tanh\Bigl(\frac{\Delta(T)}{2k_\text{B}T}\Bigr)} \biggl(\frac{l}{w}\biggr)
\label{EQUATION_MAIN_03b} 
\end{equation}
\end{subequations} 
where $L_{\text{K}}(T)$ stands for the kinetic inductance of the resonator (whose inductive length and width are represented, respectively, by the $l$ and $w$ parameters), $R_{\square}(T_\text{C})$ defines the normal-state sheet resistance of the thin-film at the critical transition, $\Delta(T) = \Delta(0) \tanh(1.74\sqrt{T_\text{C}/T-1})$ is the superconducting energy gap according to a standard BCS interpolation formula \cite{Schmidt:1997, Chauhan:2022, Nulens:2023, Powell:2025} (with $ \Delta(0) = 1.765 k_\text{B}T_\text{C}$), while $C_\text{G}$ and $L_\text{G}$ represent, respectively, the geometric capacitance and inductance of the device. In particular, $R_{\square}(T_\text{C})$ is experimentally determined through the previous DC resistance characterisation on the reference 4-wire structure (i.e. $R_{\square}(T_\text{C}) = (9.6 \pm 0.5)~\Omega/\square$), while $l$ and $w$ are taken as the length and width of the original central inductive line respectively (i.e. $l \simeq \text{\SI{1.4}{\milli\metre}}$ and $w \simeq \text{\SI{100}{\mu\metre}}$). The factor $l/w$ is meant to represent, therefore, an effective aspect ratio for the main current-carrying region. For the nanostructured resonators, this clearly turns into a simplified description, since the microwave current flows through a non-uniform network of hexagonal islands and partially etched nanotrenches, rather than through a perfectly homogeneous superconducting strip. As a consequence, the extracted parameters should be interpreted as effective quantities describing the overall resonator response. A more microscopic description would require a full electromagnetic simulation including the detailed current distribution within the nanostructured arrays, which is beyond the scope of the present work. Moreover, it is important to highlight the fact that \autoref{EQUATION_MAIN_03} is used here as a simplified effective-circuit model for the temperature calibration of the resonators. As a matter of fact, such a model includes geometrical $LC$-parameters and the temperature-dependent kinetic inductance $L_\text{K}(T)$, but it does not take into account an explicit Josephson inductance or a discrete weak-link contribution associated with the individual partially etched nanotrenches. In this sense, the effect of the nanotrenches is incorporated phenomenologically through an effective modification of the superconducting properties of each device, as reflected by the fitted parameters. Indeed, the extracted parameters should be interpreted as effective values describing the overall nanostructured resonator, rather than as microscopic parameters of each individual nanotrench. Despite this simplified description, the model provides a fitting accuracy $R^2$ larger than $\text{\text{\SI{99.9}{\percent}}}$ for all devices and is, therefore, suitable as a calibration curve for estimating the temperature sensitivity of each resonator.

In \autoref{TABLE_MAIN_01} we report the values of the three fitting parameters for such an $L_\text{K}(T)$-based model (i.e. $T_\text{C}$, $L_\text{G}$ and $C_\text{G}$), as well as the temperature sensitivity of each device derived at $\text{\SI{4.2}{\kelvin}}$ [see \autoref{MAIN_FIG_07:b}]. As it is possible to notice, an increase of such a FOM is recorded, with a reference value of $\text{\SI{5.42\pm0.03}{\mega\hertz/\kelvin}}$, obtained for the non-nanostructured device, and a maximum of $\text{\SI{62\pm1}{\mega\hertz/\kelvin}}$ registered for the $g_\text{f} = \text{\SI{350}{\nano\metre}}$ resonator. This enhancement of the temperature sensitivity, by one order of magnitude, is a direct consequence of the critical temperature shift towards lower values, for larger edge nanotrenches. Indeed, a maximum $T_\text{C}$ swing of about $\text{\SI{1.6}{\kelvin}}$ is recorded between such resonators, in agreement with the results concerning the previous DC characterisation of single-nanotrenches. This observation confirms, therefore, the initially hypothesised effect of the partially etched features, strategically placed on the outermost edges of an inductive line, on the temperature sensitivity of superconducting planar resonators, allowing to boost such an FOM by a factor $\text{\SI{10}{}}$ at $\text{\SI{4.2}{\kelvin}}$. Finally, \autoref{MAIN_FIG_07:c-e} report the temperature evolution of the quality factors for the different devices. Superimposed to the typical degradation of $Q_\text{i}$ for increasing temperatures \cite{Yu:2022, Alexander:2025}, caused by the higher density of quasiparticles $n_\text{N}(T)$ when approaching the superconducting transition $T_\text{C}$, an overall reduction of such a parameter is also observed among the investigated resonators. As a matter of fact, at $\text{\SI{4.2}{\kelvin}}$, a maximum value of $Q_\text{i} \simeq \text{8$\times$10\textsuperscript{3}}$ is recorded for the reference non-nanostructured device, while the $g_\text{f} = \text{\SI{350}{\nano\metre}}$ sensor is characterised by a minimum of $Q_\text{i} \simeq \text{1$\times$10\textsuperscript{3}}$. The partially etched Nb\textsubscript{50}Ti\textsubscript{50} regions behave, therefore, as an additional source of microwave losses to the superconducting planar resonator and introduce, as a drawback, a design trade-off with the enhancement of the temperature sensitivity.

\subsection{LHe determination of the temperature resolution}

In order to minimise the detrimental effects of the environmental noise on the performance of the nanostructured resonators, possibly deriving from the gaseous-He temperature fluctuations and the remanent magnetic field instabilities affecting our VTI \cite{Russo:2025}, the devices are tested at $\text{\SI{4.2}{\kelvin}}$ in a $\text{\SI{100}{\liter}}$ liquid helium dewar. In particular, a low-noise characterisation is carried out by means of a microwave setup relying on frequency modulation (FM) techniques \cite{Chatel:2025}, for enhancing\hfill the\hfill readout\hfill temperature\hfill resolution\hfill achieved\hfill through
\begin{figure}[H]
\includegraphics[width=\linewidth]{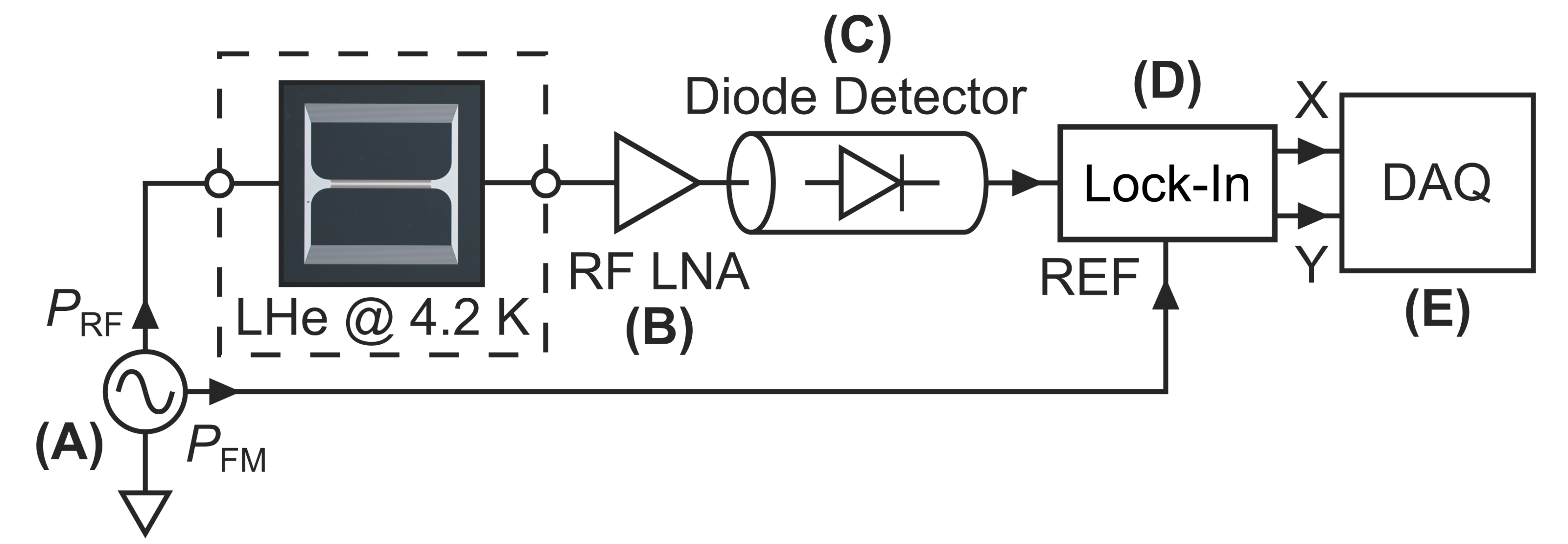}
\caption{\justifying FM low-noise instrumentation setup exploited for the characterisation of the nanostructured resonators in LHe at $\text{\SI{4.2}{\kelvin}}$. Adapted with permission from \cite{Chatel:2025}. Copyright 2025 \textit{IOP Publishing}.}
\label{MAIN_FIG_08}
\end{figure}
\noindent these sensors [see \autoref{MAIN_FIG_08}]. Electromagnetically coupled $\text{\SI{0.75}{\milli\metre}}$ far from a Cu GCPW, the resonators are excited by a microwave signal generator (\textbf{A}. \textit{Stanford Research SG384}), delivering a $\text{\SI{-40}{\dBm}}$ output power. The excitation signal is, additionally, frequency modulated, with $f_\text{FM} = \text{\SI{11}{\kilo\hertz}}$ and $\delta f_\text{FM} = \text{\SI{\pm5}{\kilo\hertz}}$ \cite{Chatel:2025}, and, subsequently, exploited as external reference by a lock-in amplifier. Once having excited the devices at the cryogenic stage, the signal is amplified through a low-noise amplifier (LNA) (\textbf{B}. \textit{Mini-Circuits ZRL-1150LN}), at room temperature.  Subsequently, a diode detector (\textbf{C}. \textit{Macom 2086-6000-13}) is exploited to perform a DC-conversion of the microwave signal. A synchronous demodulation of the $\text{\SI{11}{\kilo\hertz}}$ AC component is, then, performed by recurring to a lock-in amplification (\textbf{D}. \textit{EG\&G Instruments Model 7265}), by setting a $\text{\SI{100}{\milli\second}}$ time constant (i.e. $\text{\SI{2.5}{\hertz}}$ equivalent noise bandwidth). In the end, a DAQ (\textbf{E}. \textit{PCI-6052E}) is configured to simultaneously acquire both the in-phase (X) and quadrature (Y) components of the sensing signal with a $\text{\SI{100}{\milli\second}}$ integration time.

\begin{figure}[H]
\includegraphics[width=\linewidth]{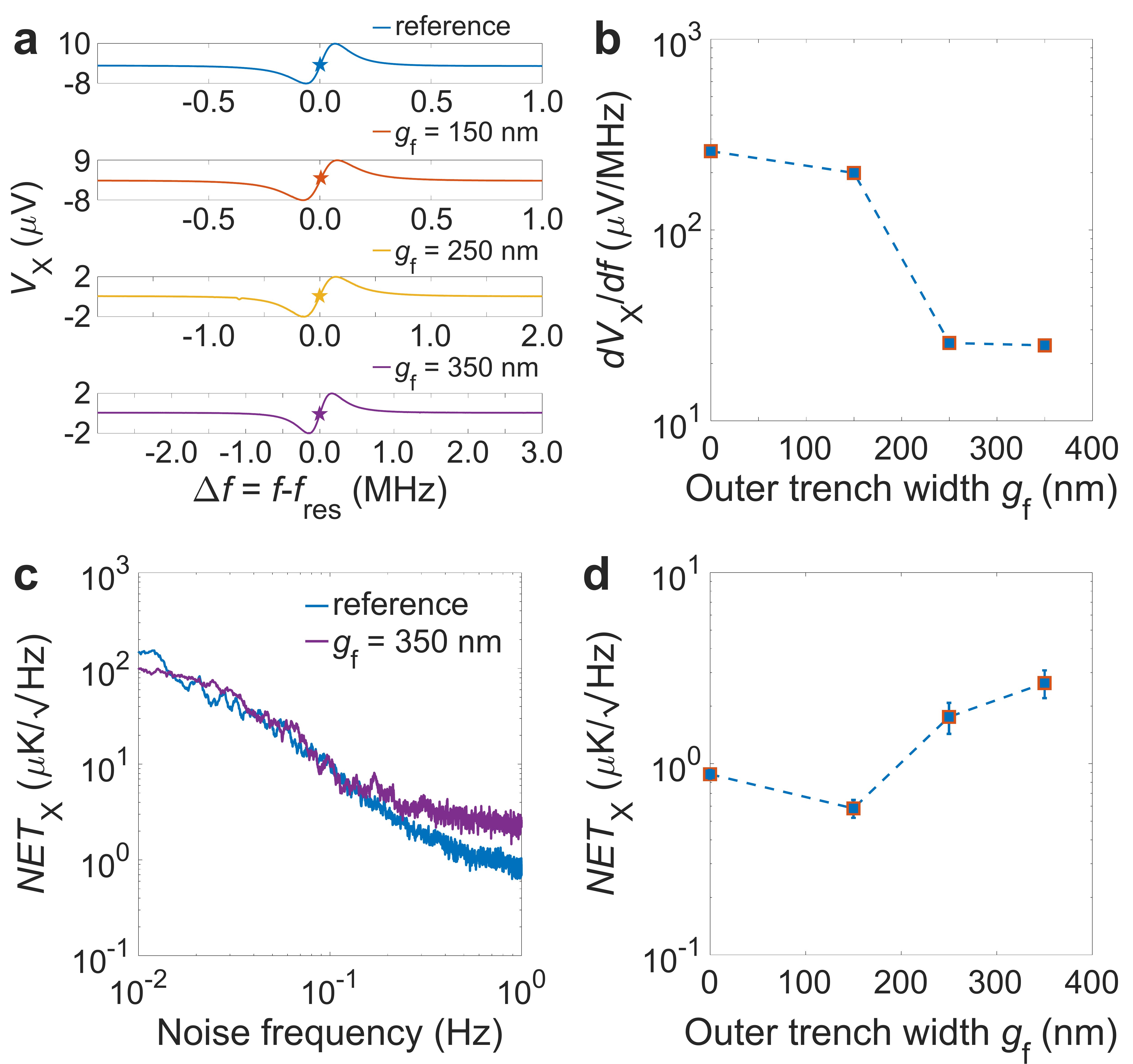}
\caption{\justifying Determination of the temperature resolution of the different nanostructured resonators at $\text{\SI{4.2}{\kelvin}}$. a) Demodulated lock-in signal, in the X component: the pentagram points stand for the condition of maximum slope, specifically located at the zero-crossing $f_\text{res}$. b) Maximum slope versus outermost nanotrench width, for the X component of the sensing signal. c) $NET$ frequency spectra for the reference and the $g_\text{f} = \text{\SI{350}{\nano\metre}}$ sensors. d) $NET$, associated to the demodulated X signal, for the different nanostructured resonators, at $\text{\SI{1}{\hertz}}$.}
\label{MAIN_FIG_09}
\makeatletter
\let\save@currentlabel\@currentlabel
\edef\@currentlabel{\save@currentlabel(a)}\label{MAIN_FIG_09:a}
\edef\@currentlabel{\save@currentlabel(b)}\label{MAIN_FIG_09:b}
\edef\@currentlabel{\save@currentlabel(c)}\label{MAIN_FIG_09:c}
\edef\@currentlabel{\save@currentlabel(d)}\label{MAIN_FIG_09:d}
\makeatother
\end{figure}

The different resonance conditions are identified by means of frequency scans, recording the lock-in output X signal [see \autoref{MAIN_FIG_09:a}]. As it is possible to notice, the effect of the Q-factor degradation, between different sensors, directly translates into a significant decrease of the zero-crossing slope $dV/df$ related to larger nanotrench widths. In particular, as reported in \autoref{MAIN_FIG_09:b}, a maximum value of about $\text{\SI{260}{\mu\volt/\mega\hertz}}$ is recorded for the reference non-nanostructured resonator (in correspondence of the resonance zero-crossing level for the in-phase X signal), while the minimum $\text{\SI{25}{\mu\volt/\mega\hertz}}$ condition is affecting the device with an outermost nanotrench width of $\text{\SI{350}{\nano\metre}}$.

Once having estimated the frequency-to-voltage conversion factors from the slope of the sensing X signals, the temperature resolution of each device can be determined by recurring to the notion of noise equivalent temperature ($NET$). By extracting the noise spectral density (NSD) of each sensing signal (in $\text{\SI{}{\volt/\hertz^{1/2}}}$) through a fast-Fourier transform (FFT) applied on $\text{\SI{2}{\hour}}$ long time-scans (with the DAQ integration time set to $\text{\SI{100}{\milli\second}}$), the associated noise spectrum can be converted into an equivalent temperature (in $\text{\SI{}{\kelvin/\hertz^{1/2}}}$) \cite{Chatel:2025}, expressed as
\begin{subequations}
\label{EQUATION_MAIN_04}
\begin{equation}
NET(f_{\text{noise}}) = \frac{NSD(f_{\text{noise}})}{R}
\label{EQUATION_MAIN_04a}
\end{equation}
\begin{equation}
R = \dfrac{dV}{dT} = \dfrac{dV}{df} \cdot \dfrac{df}{dT}
\label{EQUATION_MAIN_04b}
\end{equation}
\end{subequations}
\noindent where $R$ represents the temperature responsivity of the device (in $\text{\SI{}{\volt/\kelvin}}$). In particular, such a parameter can be obtained from the product between the temperature sensitivity $df/dT$ and the maximum slope conversion factor $dV/df$, previously estimated. The $NET$, associated to the demodulated X signal, is shown in \autoref{MAIN_FIG_09:c}, for the limiting cases of the non-nanostructured reference resonator and the $g_\text{f} = \text{\SI{350}{\nano\metre}}$ device. Although the two spectra show identical trends below $\text{\SI{100}{\milli\hertz}}$, probably induced by environmental temperature-related instabilities affecting the LHe bath \cite{Donnelly:1998, Chatel:2025}, a significant deviation appears for higher frequencies. Since $\text{\SI{4.2}{\kelvin}}$ still represents a higher temperature condition, with respect to the sub-$\text{\SI{}{\kelvin}}$ range at which atomic two-level systems (TLSs) typically limit the noise performance of Nb-based superconducting devices \cite{Lindstrom:2009, Burnett:2013, Muller:2019}, the observed discrepancy is, most likely, induced by the electronic noise associated to the different density of quasiparticles for each resonator \cite{Lindstrom:2011}. In particular, the ideal thermal noise $\sqrt{4k_\text{B}TR}$ associated to the microwave resistance of the resonators can be estimated to be in the order of few $\text{\SI{}{\pico\volt/\hertz^{1/2}}}$ at $\text{\SI{4.2}{\kelvin}}$, resulting in a limiting resolution of about $\text{\SI{20}{}-\SI{60}{\nano\kelvin/\hertz^{1/2}}}$, after applying the responsivity voltage-to-temperature conversion from \autoref{EQUATION_MAIN_04b}. More in detail, the $NET$ obtained at $\text{\SI{4.2}{\kelvin}}$ and for a $\text{\SI{1}{\hertz}}$ noise frequency, turns out to experience a deterioration for an increasing value of the edge nanotrench width [see \autoref{MAIN_FIG_09:d}], reaching a value as large as $\text{\SI{2.6\pm0.4}{\mu\kelvin/\hertz^{1/2}}}$ for the limiting $g_\text{f} = \text{\SI{350}{\nano\metre}}$ case. Even though the presence of the thinned superconducting regions induces a worsening of the temperature resolution, by almost a factor $\text{\SI{3}{}}$ with respect to the reference resonator, such devices still offer a temperature resolution in the order of few $\text{\SI{}{\mu\kelvin}}$, still outperforming the cryogenic temperature sensors available on the market by more than one order of magnitude.

\begin{table*}[t!]
\begin{threeparttable}
\caption{\justifying Performance comparison across different sensors for superconducting cryogenic thermometry. The midrule is intended to separate the temperature performance associated to devices from previous literature studies from the results reported in this work, while the uncertainties associated to each parameter have been intentionally omitted, for the sake of clarity and readability. Successive repeated entries in columns are indicated with a ditto mark ("). Missing data is indicated with a dash (-).}
\label{TABLE_MAIN_02}
\renewcommand{\arraystretch}{1.25}
\begin{tabular*}{\linewidth}{@{\hspace{3pt}\extracolsep{\fill}} ccccccc @{\hspace{3pt}}}
\toprule
Reference & Material & Type of device & \makecell{Operating temperature\\$\text{(\SI{}{\kelvin})}$} & \makecell{$df_\text{res}/dT$\tnote{a}\\$\text{(\SI{}{\hertz/\kelvin})}$} & \makecell{$Q_\text{i}$\tnote{a}\\$\bigl(\text{$\times$10\textsuperscript{3}}\bigr)$} & \makecell{$NET$\tnote{a}\\$\bigl(\text{\SI{}{\kelvin/\hertz^{1/2}}}\bigr)$}\\
\midrule
\cite{Faivre:2014} & Ti & Josephson junction & $\sim0.3$ & - & - & $2\times10^{-6}$\\
\cite{Lvov:2025} & Al & Transmon qubit & $<0.2$ & - & - & $3\times10^{-1}$\\
\cite{Noah:2024} & - & \makecell{Superconducting phase\\transition sensor} & $<1.2$ & - & - & $1\times10^{-2}$\\
\cite{Wheeler:2020}& Nb & TLS-resonator & $<1.0$ & $1\times10^{4}$ & $150$ & $5\times10^{-5}$\\
\cite{Chatel:2025} & \makecell{Nb\textsubscript{50}Ti\textsubscript{50}\\($t = \text{\SI{100}{\nano\metre}}$)} & $L_\text{K}(T)$-resonator & $\sim4.2$ & $2\times10^{6}$ & $26$ & $5\times10^{-7}$\\
" & \makecell{Nb\textsubscript{50}Ti\textsubscript{50}\\($t = \text{\SI{20}{\nano\metre}}$)} & " & " & $19\times10^{6}$ & $3$ & $7\times10^{-6}$\\
\midrule
This work& \makecell{Nb\textsubscript{50}Ti\textsubscript{50}\\(reference)} & $L_\text{K}(T)$-resonator & $\sim4.2$ & $5\times10^{6}$ & $8$ & $9\times10^{-7}$\\
" & \makecell{Nb\textsubscript{50}Ti\textsubscript{50}\\($g_\text{f} = \text{\SI{350}{\nano\metre}}$)} & \makecell{Nanostructured\\$L_\text{K}(T)$-resonator} & $"$ & $62\times10^{6}$ & $1$ & $3\times10^{-6}$\\
\bottomrule
\end{tabular*}
\begin{tablenotes}
\item[a] \footnotesize{Parameters estimated for the devices characterised at the corresponding operating temperature conditions.}
\end{tablenotes}
\end{threeparttable}
\end{table*}

Nevertheless, different design strategies might be adopted in future researches to mitigate the impact of the proposed surface nanostructuring approach on the $Q$-factor, quasiparticle density and, consequently, temperature resolution of such devices. As a matter of fact, electromagnetic simulations of the microwave superconducting current density should be exploited to investigate the possibility to strategically place the nanotrenches only in regions where they efficiently contribute to the kinetic-inductance response, while avoiding unnecessary lossy areas. Such an optimisation would not remove the intrinsic trade-off between sensitivity and resolution, but could help limit the degradation of the $NET$ while exploiting the enhanced responsivity of the nanostructured resonators. Additionally, alternative surface-modification techniques could also be investigated. For instance, greyscale lithography \cite{Stavenga:2023, Khonina:2024, Erbas:2024} could be used to gradually taper down the thickness of the main inductive line along its cross section, with the aim of realising a more uniform and discontinuity-free device. Such an approach would still rely on the increase of kinetic inductance in locally thinner superconducting films, while potentially reducing the additional microwave losses introduced by abrupt etched nanotrenches.
\subsection{Performance benchmark against literature}
In order to quantitatively assess the efficacy of the proposed nanostructuring mechanism, the devices developed alongside this work have to be compared with previous superconducting thermometers from the literature. In this perspective, \autoref{TABLE_MAIN_02} clearly shows that the temperature sensitivities $df_\text{res}/dT$ obtained in the present work are among the highest reported for superconducting microwave thermometers operating around LHe temperature. As a matter of fact, such an enhancement, by approximately one order of magnitude (i.e. up to $\text{\SI{62\pm1}{\mega\hertz/\kelvin}}$), originates from the deliberate suppression of superconductivity within the nanostructured regions, shifting the effective transition closer to the operating temperature. However, the comparison also highlights that an increased sensitivity alone is not sufficient to guarantee an improvement of the overall thermometric performance, as the associated reduction of $Q_\text{i}$ and the increasing density of quasiparticles negatively impact the ultimate $NET$. Nevertheless, in contrast to previous approaches relying on different physical phenomena \cite{Wheeler:2020} or the reduction of the global film thickness \cite{Chatel:2025}, the proposed nanostructuring strategy exploits local modifications of the superconducting properties within selected regions of the sensor, introducing an additional design parameter that can be optimised independently from the overall resonator geometry or the specific material platform.

\section{Conclusion}
Through this work, we have demonstrated the possibility to tune, and enhance, the temperature sensitivity of superconducting microwave resonators, by strategically nanostructuring the surface of such thin-film devices. In such a perspective, a set of different $\text{\SI{100}{\nano\metre}}$ thick Nb\textsubscript{50}Ti\textsubscript{50} superconducting microwave resonators has been designed, fabricated and characterised below $\text{\SI{10}{\kelvin}}$. More technically, a high-resolution combination of e-beam lithography and fluorine-based plasma etching has been exploited to partially etch the superconducting thin-film and, therefore, pattern different $\text{\SI{20}{\nano\metre}}$ thick regions on the main inductive line of a microwave resonator.

First, we have preliminarily characterised the electronic transport properties of single-nanotrenches, with widths of $\text{\SI{50}{}}-\text{\SI{350}{\nano\metre}}$, by means of 4-wire DC tests. The resulting $I$/$V$ curves show a net effect of the extension of each feature on the superconducting critical transition. In particular, a gradual $\text{\SI{1.5}{\kelvin}}$ decrease of the superconducting critical temperature $T_\text{C}$ is recorded for increasing nanotrench widths. Such a trend has also been successively observed on the temperature response of the nanostructured microwave resonators, with Q-factors generally larger than $\text{10\textsuperscript{3}}$ at $\text{\SI{4.2}{\kelvin}}$. In particular, the $\text{\SI{1.5}{\kelvin}}$ variation of $T_\text{C}$, associated to the different nanotrenches distributions, directly translates into a larger bending of the $f_\text{res}(T)$ calibration curve. In this framework, we have successfully recorded a factor $\text{\SI{10}{}}$ increase of the temperature sensitivity, at $\text{\SI{4.2}{\kelvin}}$, to a value as large $\text{\SI{62\pm1}{\mega\hertz/\kelvin}}$ for the $g_\text{f} = \text{\SI{350}{\nano\metre}}$ resonator. Nevertheless, the Q-factor degradation, associated to larger microwave losses induced by the presence of the partially etched trenches, is found to impose a performance trade-off when considering the temperature resolution of such devices. A low-noise FM-based characterisation has also been carried out in liquid He at $\text{\SI{4.2}{\kelvin}}$, revealing that, as a consequence of the increased losses associated to the higher density of quasiparticles, the ultimate temperature resolution, at $\text{\SI{1}{\hertz}}$, experiences a deterioration by a factor $\text{\SI{3}{}}$ with respect to the reference condition, with the highest value of $\text{\SI{2.6\pm0.4}{\mu\kelvin/\hertz^{1/2}}}$ achieved for the $g_\text{f} = \text{\SI{350}{\nano\metre}}$ resonant. Nevertheless, such an FOM is still better than sensors typically exploited in cryogenic applications, by at least one order of magnitude, and in agreement with previous literature studies on similar non-nanostructured superconducting devices \cite{Chatel:2025}.

In conclusion, the surface nanostructuring of thin-film superconducting microwave resonators investigated in this work provides an additional design tool for tailoring the temperature-dependent frequency response of such devices. The results demonstrate that locally thinning the superconducting film into arrays of nanotrenches can effectively enhance the temperature sensitivity around a target operating condition. However, at the same time, the observed decrease of the quality factor highlights an intrinsic trade-off between the enhanced temperature sensitivity and a degraded $NET$, limited to the few-$\text{\SI{}{\mu\kelvin/\hertz^{1/2}}}$ range (at $\text{\SI{1}{\hertz}}$ and $\text{\SI{4.2}{\kelvin}}$). Future optimisations should, therefore, focus on mitigating the RF losses introduced by the nanostructures, while preserving the enhancement of the $T$-sensitivity offered by this approach. Moreover, besides the specific Nb\textsubscript{50}Ti\textsubscript{50} platform investigated here, the proposed nanostructuring strategy may provide a versatile route for tailoring the thermometric response of superconducting microwave resonators. Such devices could find application in multiplexed cryogenic thermometry, on-chip thermal monitoring of superconducting and quantum circuits and low-temperature studies of thermal transport and quantum thermodynamics. More generally, the ability to locally engineer the superconducting properties of a resonator through the nanoscale modulation of the thin-film thickness introduces an additional design degree of freedom for the development of the next-generation cryogenic thermometers.

%
%

\ack{The authors would like to express their gratitude to the EPFL Center of Micro/Nanotechnology (CMi), and, specifically, to Cyrille Hibert, Zdenek Benes, Julien Dorsaz, Niccolò Piacentini, Joffrey Pernollet, Makhlad Chahid and Adrien Toros, for the facility support provided during the microfabrication processing. We would also like to thank Mounir Driss Mensi, for the help with the XPS analysis, as well as Pasquale Scarlino, Franco De Palma and Reza Farsi, for the fruitful discussions.}

\funding{This work has been funded by the Swiss National Science Foundation (SNSF), through the AMBIZIONE program (Project "Cryogenic Thermometry based on Superconducting Microwave Resonators", grant agreement No. PZ00P2\_193361).}

\roles{
\textbf{A. Chatel}: Data curation (lead); Formal analysis (lead); Investigation (lead); Methodology (lead); Software (lead); Visualization (lead); Writing - original draft (lead); Writing - review \& editing (equal);
\textbf{R. Russo}: Data curation (supporting); Formal analysis (supporting); Investigation (supporting); Methodology (equal); Software (equal); Writing - review \& editing (equal);
\textbf{S. A. Hashemi}: Formal analysis (supporting); Investigation (supporting); Methodology (supporting); Software (supporting); Writing - review \& editing (equal);
\textbf{P. Torrev-Vila}: Formal analysis (supporting); Investigation (supporting); Software (supporting); Writing - review \& editing (equal);
\textbf{J. Brügger}: Resources (equal); Supervision (equal); Validation (supporting); Writing - review \& editing (equal);
\textbf{G. Boero}: Conceptualization (equal); Data curation (equal); Formal analysis (equal); Funding acquisition (equal); Investigation (equal); Methodology (equal); Project administration (lead); Resources (equal); Software (equal); Supervision (lead); Validation (lead); Writing - original draft (equal); Writing - review \& editing (lead);
\textbf{H. Furci}: Conceptualization (lead); Data curation (equal); Formal analysis (equal); Funding acquisition (lead); Investigation (equal); Methodology (equal); Project administration (lead); Resources (lead); Software (equal); Supervision (lead); Validation (equal); Writing - original draft (equal); Writing - review \& editing (lead).
}

\data{The data that support the findings of this study are available upon reasonable request from the authors.}

\suppdata{The \textcolor{blue}{supplementary data} contains additional information about the following topics:
\vspace{-6pt}
\begin{enumerate}
\setlength{\itemsep}{-4pt}
\item[S1.] Technical details concerning the micro and nanofabrication process flow;\\
\item[S2.] Microwave simulation of the reference resonator;\\
\item[S3.] Preliminary characterisation of the nanostructured resonators in liquid He;\\
\item[S4.] AFM scans of the different single-nanotrench 4-wire structures;\\
\item[S5.] Typical $I$/$V$ dataset for a sweep of the cryogenic temperature $T$.\\
\end{enumerate}}

\section*{Conflict of interest}
The authors have no conflicts to disclose.

\section*{References}
\printbibliography[heading=none]

\end{multicols*}

\end{document}


\begin{center}
\LARGE{Thin-film nanostructuring of NbTi superconducting resonators for enhanced temperature sensitivity in cryogenic thermometry: Supplementary data}\\
\vspace{4mm}
\Large{A. Chatel,\textsuperscript{1-2} R. Russo,\textsuperscript{1-2} S. A. Hashemi,\textsuperscript{1} P. Torres-Vila,\textsuperscript{1}\\ J. Brugger,\textsuperscript{1} G. Boero\textsuperscript{1-2} and H. Furci\textsuperscript{1}}\\
\vspace{4mm}
\large{\textsuperscript{1}Microsystems Laboratory, EPFL, 1015 Lausanne, Switzerland}\\
\large{\textsuperscript{2}Center for Quantum Science and Engineeing, EPFL, 1015 Lausanne, Switzerland}\\
\end{center}
\vspace{4mm}
\normalsize

\section{Microwave simulation of the reference non-nanostructured resonator}
\label{SUPPL_MAT_01}

In this section, we describe the details concerning the electromagnetic Finite Element Method (FEM) simulation performed with the aid of COMSOL Multiphysics\textsuperscript{\textregistered} software. More specifically, such a study is carried out with the aim of assessing the antisymmetric $B$-field configuration necessary to couple the device with a Cu Grounded Coplanar Waveguide (GCPW) and preliminarily estimating its specific resonance frequency.\\

\begin{figure*}[h!]
\includegraphics[width=\textwidth]{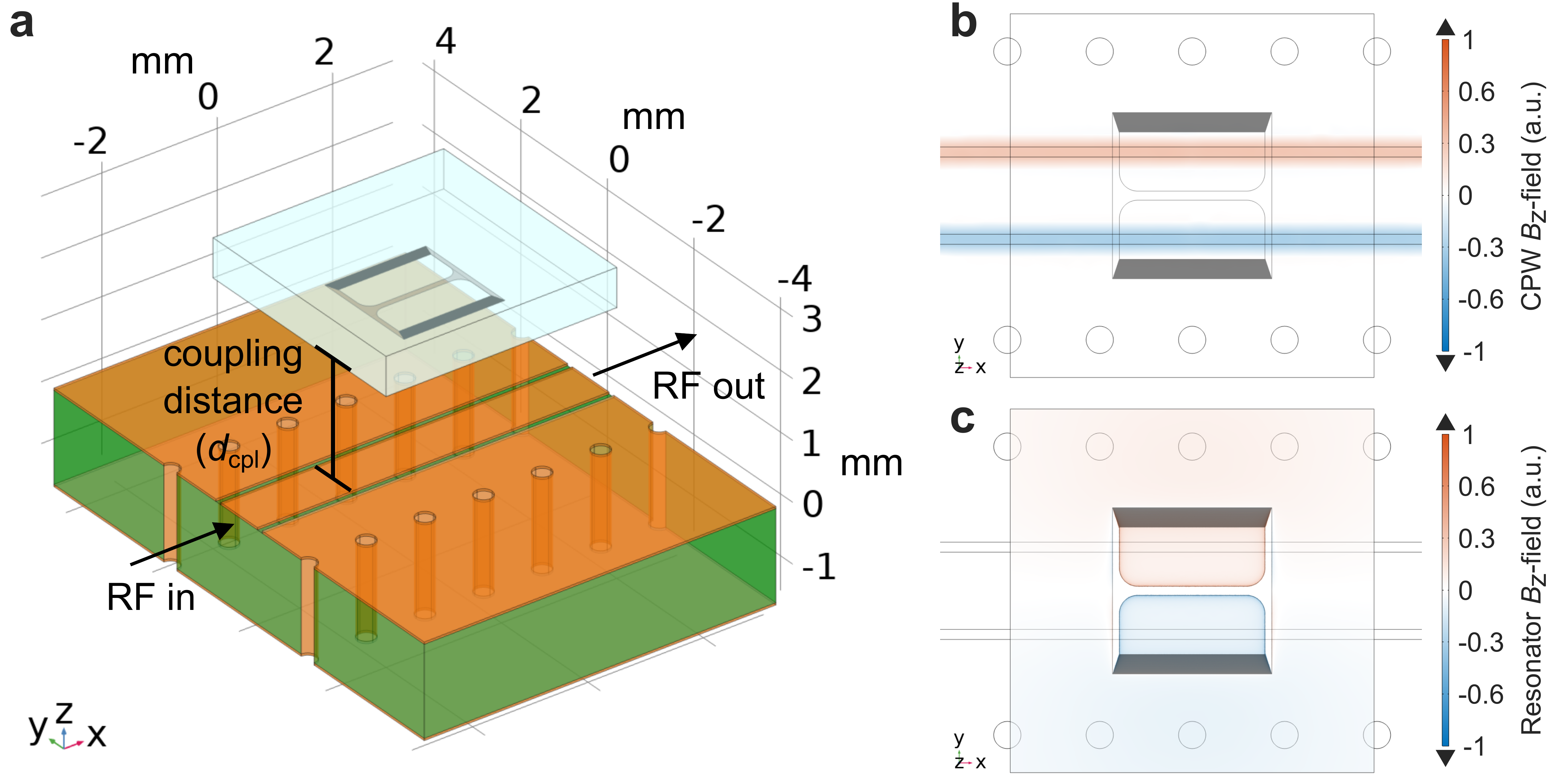}
\caption{COMSOL Multiphysics\textsuperscript{\textregistered} microwave simulation of the reference resonator, coupled to a Cu GCPW. a) Schematic of the simulation setup, showing the reference PEC resonator electromagnetically coupled, via geometric proximity, to a Cu excitation GCPW. b) Antisymmetric $B$-field distribution of the quasi-TEM mode associated to the GCPW transmission line. c) Antisymmetric $B$-field distribution of the fundamental resonance eigenmode ($f_\text{res} \sim \text{\SI{1.3}{\giga\hertz}}$).}
\label{SUPPL_FIG_01}
\makeatletter
\let\save@currentlabel\@currentlabel
\edef\@currentlabel{\save@currentlabel(a)}\label{SUPPL_FIG_01:a}
\edef\@currentlabel{\save@currentlabel(b)}\label{SUPPL_FIG_01:b}
\edef\@currentlabel{\save@currentlabel(c)}\label{SUPPL_FIG_01:c}
\edef\@currentlabel{\save@currentlabel(b-c)}\label{SUPPL_FIG_01:b-c}
\makeatother
\end{figure*}

The implemented 2-port simulation setup is depicted in \autoref{SUPPL_FIG_01:a}, showing the transmission of the excitation microwave power along the GCPW. In particular, the characteristic quasi-TEM (Transverse Electro-Magnetic) mode of the excitation line is enforced by defining the ports as boundary conditions for this specific electromagnetic field distribution. Indeed, this approach turns out to be particularly useful for such an eigenmode study, as it allows to limit the number of the resonator's eigenfrequencies to the few configurations that can couple with the coplanar line \cite{Simons:2001}.\\
The Printed Circuit Board (PCB) exploited to route the microwave electromagnetic field can be modelled as a $\text{\SI{34}{\mu\metre}}$ thick Cu GCPW, on top of a $\text{\SI{1.5}{\milli\metre}}$ thick ROGERS 4003C\texttrademark{} laminate ($\epsilon_\text{r} \simeq 3.38$ and $\tan\delta \simeq 0.0027$), whose central signal line is $\text{\SI{850}{\mu\metre}}$ wide and $\text{\SI{110}{\mu\metre}}$ laterally spaced from the external ground plane. Additionally, the reference resonator (i.e. without any surface nanostructures) is placed face-down, $\text{\SI{2}{\milli\metre}}$ far from the Cu GCPW, in order to  realise an inductive coupling, via geometric proximity.\\
Because of the very high aspect-ratio affecting the device, the superconducting thin-film is simply modelled as a 2D Perfect Electric Conductor (PEC), ideally showing an infinite conductivity. Although the microwave properties of a superconducting thin-film are not represented by such a model, its simplicity turns out to provide a sufficient accuracy, and enough computational efficiency, to determine the desired resonance frequency to be in the order of  $\text{\SI{1.3}{\giga\hertz}}$, consistent with the approximated value estimated in the main text. The confirmation of the proper inductive coupling is reported in \autoref{SUPPL_FIG_01:b-c}, showing the antisymmetric $B$-field distributions for both the GCPW and the PEC resonator. Finally, the port matching is also determined, extracting the Cu line characteristic impedance to be about $Z_\text{C} \simeq \text{\SI{48}{\ohm}}$. Such a value further provides an effective validation of the designed microwave $\text{\SI{50}{\ohm}}$-matched GCPW.

\section{Technical details concerning the micro-nanofabrication process flow}
\label{SUPPL_MAT_02}

The aim of this section if to discuss about the micro-nanofabrication of the nanostructured resonators, defining the process flow exploited, first, to partially etch weakened superconducting nanotrenches into a Nb\textsubscript{50}Ti\textsubscript{50} thin-film and, successively, to pattern the microwave resonator geometry all around.\\

\begin{figure*}[t!]
\includegraphics[width=\textwidth]{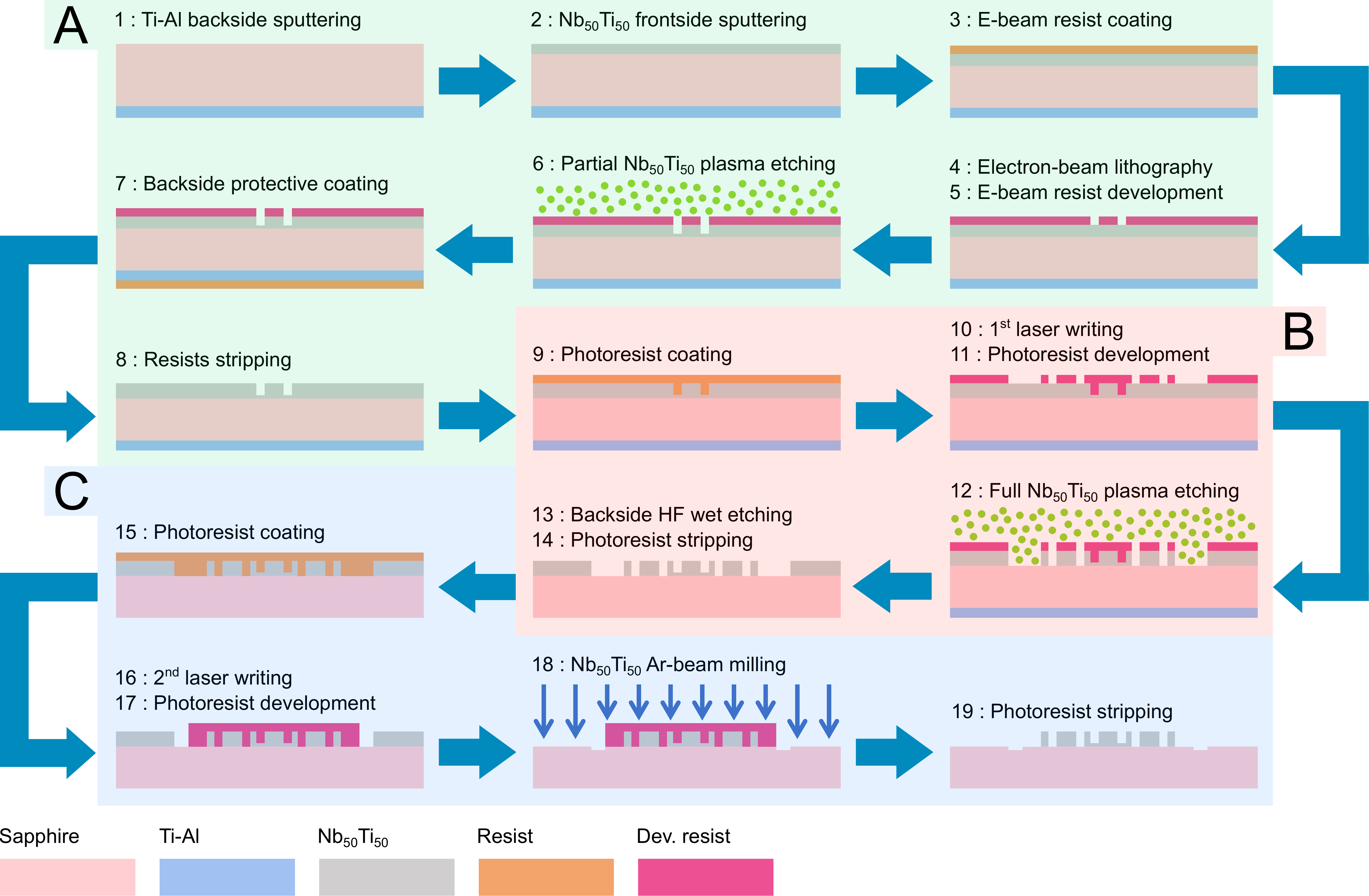}
\caption{Micro-nanofabrication process flow to pattern arrays of partially etched nanotrenches on Nb\textsubscript{50}Ti\textsubscript{50} thin-films, DC sputtered on top of sapphire wafers. Sequence A is dedicated to the generation of the partially etched nanotrenches, while sequences B and C are exploited to pattern the planar resonating structure, as described in \cite{Chatel:2025}.}
\label{SUPPL_FIG_02a}
\end{figure*}

\autoref{SUPPL_FIG_02a} outlines the different steps exploited to transfer arrays of nanostructures on Nb\textsubscript{50}Ti\textsubscript{50} thin-films, deposited on top of C-plane sapphire wafers $\text{\SI{650}{\mu\metre}}$ thick. The process flow consists of three sequences, A, B and C, each one resulting from a combination of a lithographic patterning step and a dry etching one. In particular, the array of weakened-superconducting regions is realised by partially etching the Nb\textsubscript{50}Ti\textsubscript{50} thin-film, in order to locally create nanotrenches of lower thickness. As a matter of fact, since several studies have been reporting the superconducting properties of a thin-film to degrade with a reduction of the material thickness \cite{Minhaj:1994, Takeda:1999, Ilin:2004, Gubin:2005, Inomata:2009, Sibanda:2022, Bretz-Sullivan:2022, Zhu:2024, Chatel:2025}, this approach should induce the local suppression of the Nb\textsubscript{50}Ti\textsubscript{50} order parameter and, therefore, introduce a weakened-superconducting feature in the form of a partially etched nanotrench.\\
It is important to highlight the fact that the nanostructuring of a metal-superconductor bilayer (e.g. Pt-Nb\textsubscript{50}Ti\textsubscript{50} or Al-Nb\textsubscript{50}Ti\textsubscript{50}) has also been initially considered, bridging superconducting islands by means of the proximity effect induced by a normal metal film underneath \cite{Holm:1932, Meissner:1960, Kircher:1968, Clarke:1968, Likharev:1979}. However, this approach leads to the formation of an interface which may significantly deteriorate the quality factor of the resonators (i.e. by dramatically increasing the amount of RF losses), as a consequence of an excessive presence of surface defects and the injection of additional quasiparticles, from the normal metal, into the superconducting layer. For this reason, a simpler, cleaner and interface-free approach is adopted, solely relying on the nanostructuring of a single superconducting thin-film.\\

Considering, initially, sequence A (dedicated to the transfer of the partially etched nanotrenches), the first two steps involve the room-temperature DC sputtering of a $\text{\SI{10}{\nano\metre}}$-$\text{\SI{100}{\nano\metre}}$ Ti-Al backside layer and a $\text{\SI{100}{\nano\metre}}$ thick superconducting Nb\textsubscript{50}Ti\textsubscript{50} thin-film. In particular, the Ti-Al backside layer is necessary to implement an electrostatic clamping of the wafer when performing the Nb\textsubscript{50}Ti\textsubscript{50} plasma etching.\\
After the deposition of the superconducting thin-film, a combination of Electron-Beam Lithography (EBL) and SF\textsubscript{6}/CHF\textsubscript{3}-based plasma etching is exploited to pattern an array of nanotrenches. More precisely, such features, partially etched inside the Nb\textsubscript{50}Ti\textsubscript{50} thin-film, are meant to leave a residual $\text{\SI{20}{}}$-$\text{\SI{30}{\nano\metre}}$ thick superconducting material, bridging together the thicker islands. The precise control of the nanotrench depth is achieved by means of the relatively slow etching process, characterised by a Nb\textsubscript{50}Ti\textsubscript{50} etch rate of about $\text{\SI{6.7}{\angstrom/\second}}$.\\
The e-beam resist is, then, stripped through a combination of dry and wet processes, including a first dip into a $\text{\SI{1}{\percent}}$ diluted HF bath (to remove the fluorinated sidewalls formed during the plasma etching step), with the backside bilayer being protected against the aggressive action of such an acid by means of preliminary photoresist coating.\\

The successive role of sequences B and C is to define the planar structure of the resonator. In particular, an optimised high-resolution combination of Direct Laser Writing (DLW) and the same previous fluorine-based plasma etching is, first, exploited to locally pattern the geometry of the resonator in $\text{\SI{2.4}{\milli\metre}} \times \text{\SI{2.4}{\milli\metre}}$ squared openings. As described in \cite{Chatel:2025}, the minimisation of the etching area is intentional, in order to reduce any thermal stress originating from the exposure of large portions of the metallic film to the plasma, which induce, as a consequence, the detachment of the wafer from the electrostatic clamping mechanism of the etching machine.\\
Finally, the remaining Nb\textsubscript{50}Ti\textsubscript{50} film, surrounding the resonator, is removed by means of a coarser combination of DLW and Ar-beam milling, whose requirement for a photoresist reflow induce a loss of patterning accuracy and resolution with respect to sequence B. Nevertheless, the clamping mechanism of our milling machine is purely mechanical, and, thus, it turns out to be an effective tool for completing the removal of the metallic thin-film in excess.\\

\begin{figure*}[t!]
\includegraphics[width=\textwidth]{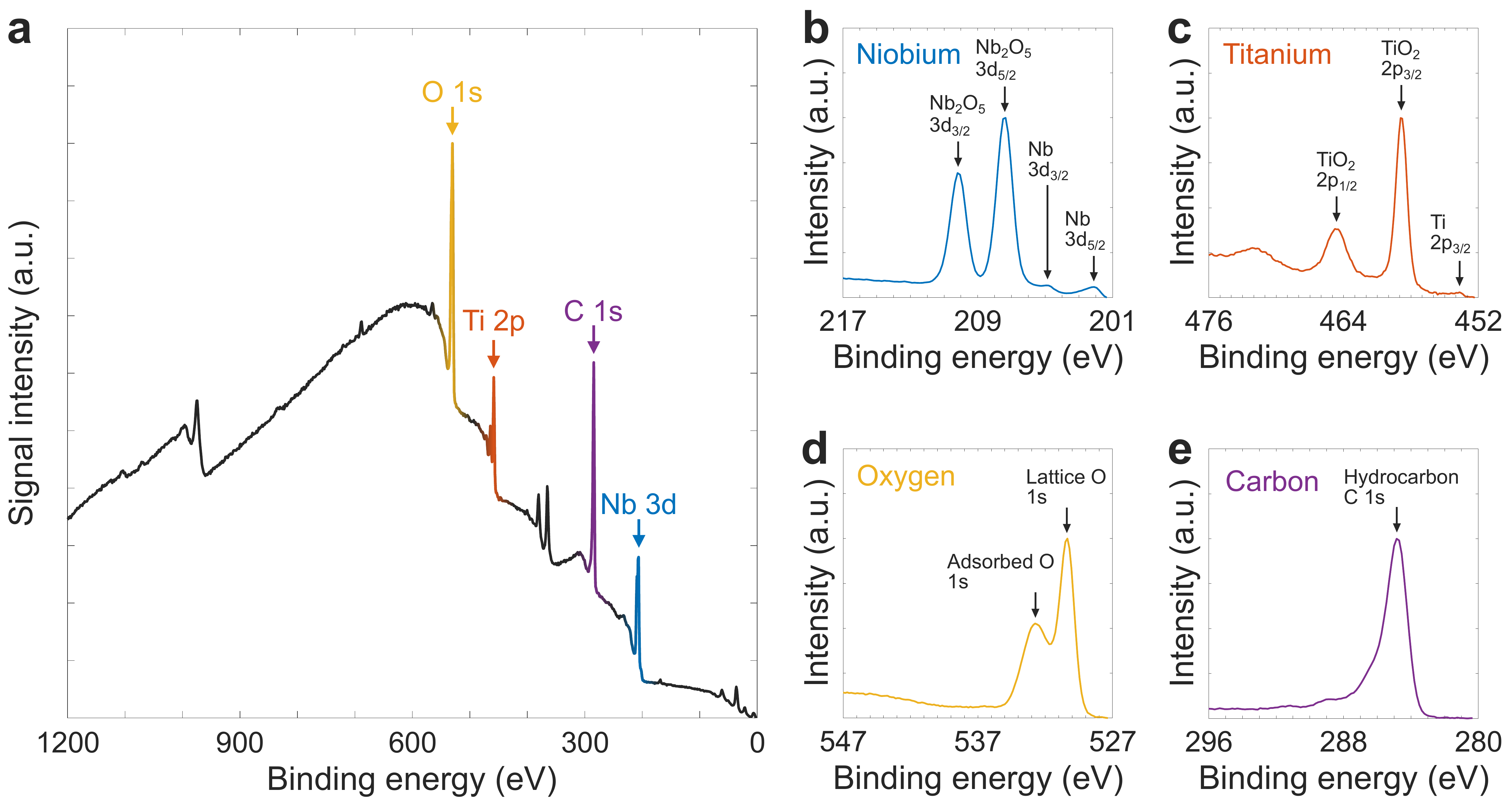}
\caption{XPS characterisation of the processed Nb\textsubscript{50}Ti\textsubscript{50} thin film. a) Survey spectrum over the whole scanning energy range, showing the presence of the characteristic Nb, Ti, O and C peaks. b) High-resolution Nb 3d spectrum, dominated by Nb\textsubscript{2}O\textsubscript{5} contributions and characterised by weaker low-energy metallic Nb components. c) High-resolution Ti 2p spectrum, dominated by TiO\textsubscript{2} contributions and characterised by a weaker low-energy metallic Ti component. d) High-resolution O 1s spectrum, showing a dominant metal-oxide contribution and a higher-energy component attributed to hydroxyl groups and adsorbed oxygen. e) High-resolution C 1s spectrum, dominated by the hydrocarbon C-C/C-H component, used for energy calibration at $\text{\SI{284.8}{\electronvolt}}$.}
\label{SUPPL_FIG_02b}
\makeatletter
\let\save@currentlabel\@currentlabel
\edef\@currentlabel{\save@currentlabel(a)}\label{SUPPL_FIG_02b:a}
\edef\@currentlabel{\save@currentlabel(b)}\label{SUPPL_FIG_02b:b}
\edef\@currentlabel{\save@currentlabel(c)}\label{SUPPL_FIG_02b:c}
\edef\@currentlabel{\save@currentlabel(d)}\label{SUPPL_FIG_02b:d}
\edef\@currentlabel{\save@currentlabel(e)}\label{SUPPL_FIG_02b:e}
\edef\@currentlabel{\save@currentlabel(b-c)}\label{SUPPL_FIG_02b:b-c}
\makeatother
\end{figure*}

In addition to the SEM and AFM morphological characterisation discussed in the main manuscript, an X-ray photoelectron spectroscopy (XPS) analysis is carried out in order to provide chemical information on the processed Nb\textsubscript{50}Ti\textsubscript{50} thin-film. In particular, such measurements are performed on an \textit{Axis Supra} spectrometer from \textit{Kratos Analytical Ltd}, exploiting the monochromated Al K$\alpha$ X-ray line of an aluminium anode. High-resolution scans are acquired with an analyser pass energy of $\text{\SI{40}{\electronvolt}}$ and an energy step size of $\text{\SI{0.15}{\electronvolt}}$. Being the sample electrically insulated, charge neutralisation procedure is achieved by exploiting a low-energy electron flood gun. Moreover, the scale of the binding energy is calibrated by referencing the C 1s C-C/C-H component, from adventitious carbon, to $\text{\SI{284.8}{\electronvolt}}$.\\
As shown in \autoref{SUPPL_FIG_02b:a}, the XPS survey spectrum confirms the presence of the characteristic peaks related to Nb, Ti, O and C. High-resolution spectra are also acquired, focusing on the different spectral regions associated to these latter elements. In particular, as reported in \autoref{SUPPL_FIG_02b:b-c}, Nb 3d and Ti 2p spectra are well-distinguishable and dominated, respectively, by Nb\textsubscript{2}O\textsubscript{5} ($E_\text{3d\textsubscript{5/2}} \simeq \text{\SI{207.4}{\electronvolt}}$) and TiO\textsubscript{2} ($E_\text{2p\textsubscript{3/2}} \simeq \text{\SI{458.9}{\electronvolt}}$) oxidised contributions, consistently with the formation of a native Nb/Ti oxide layer on the top surface directly exposed to air. Nevertheless, weaker low-energy contributions associated with metallic Nb ($E_\text{3d\textsubscript{5/2}} \simeq \text{\SI{202.3}{\electronvolt}}$) and Ti ($E_\text{2p\textsubscript{3/2}} \simeq \text{\SI{453.7}{\electronvolt}}$) are also detected, indicating the underlying alloy can still be probed through the native oxide. In particular, the observed values are found to be consistent with the ones reported by previous literature works on NbTi-based superconducting thin films \cite{Jouve:1996, Yemane:2017}. The O 1s spectrum further confirms the highly oxidised nature of the surface [see \autoref{SUPPL_FIG_02b:d}], while the one associated to the C 1s seems dominated by the hydrocarbon C-C/C-H component, used for the calibration of the binding energy, as shown in \autoref{SUPPL_FIG_02b:e}.\\
Moreover, a semi-quantitative analysis performed by the XPS facility on the metallic Nb and Ti contributions provides a Nb:Ti relative atomic percentage of about 55:45, in reasonable agreement with the nominal Nb\textsubscript{50}Ti\textsubscript{50} alloy composition of the sputtering target. However, due to the surface sensitivity of the XPS technique and the presence of an important native oxide layer, this value should be considered as an approximated surface-sensitive estimation rather than a precise measurement of the thin-film stoichiometry.

\section{Preliminary characterisation of the nanostructured resonators in liquid He}
\label{SUPPL_MAT_03}

The aim of this section is to describe the results concerning the preliminary RF tests carried out in liquid He, at $\text{\SI{4.2}{\kelvin}}$, in order to identify an ideal excitation power for all the different aforementioned resonators. In such a perspective, an experimental search for an optimal coupling condition is carried out, by varying the distance between the reference sensor and the Cu GCPW.

\subsection{Determination of the excitation power level}
\label{SUPPL_MAT_03_a}

In order to identify a suitable microwave power level $P_\text{RF}$, for exciting the devices under test without triggering any non-linear behaviour, either induced by quasiparticles self-heating or by self-Kerr non-linearities \cite{Thomas:2020}, each resonator is characterised in Liquid He (LHe), at a fixed $\text{\SI{0.75}{\milli\metre}}$ coupling distance. Such an electromagnetic coupling via geometric proximity is achieved by 3D-printing a Polyactide (PLA) support able to suspend the chips, face-down, on top of the PCB and align them with the excitation Cu GCPW.\\

\begin{figure*}[t!]
\includegraphics[width=\textwidth]{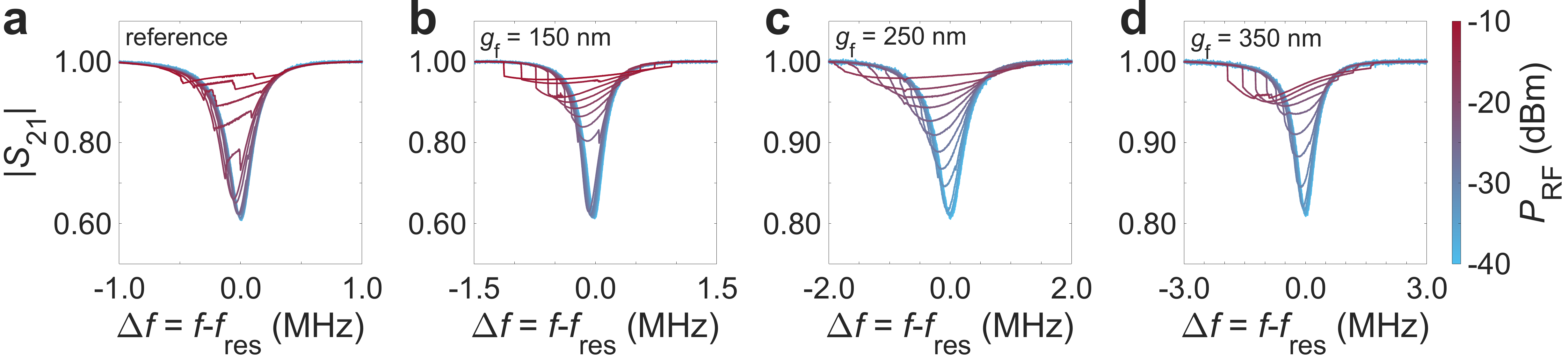}
\caption{Effect of the excitation microwave power $P_\text{RF}$ on the different nanostructured devices, at $\text{\SI{4.2}{\kelvin}}$ in LHe. a) Frequency response for the reference resonator ($f_\text{res} = \text{\SI{1.3646}{\giga\hertz}}$). b) Frequency response for the $g_\text{f} = \text{\SI{150}{\nano\metre}}$ resonator ($f_\text{res} = \text{\SI{1.3551}{\giga\hertz}}$). c) Frequency response for the $g_\text{f} = \text{\SI{250}{\nano\metre}}$ resonator ($f_\text{res} = \text{\SI{1.3423}{\giga\hertz}}$). d) Frequency response for the $g_\text{f} = \text{\SI{350}{\nano\metre}}$ resonator ($f_\text{res} = \text{\SI{1.3257}{\giga\hertz}}$).}
\label{SUPPL_FIG_03_a}
\end{figure*}

A Vector Network Analyser (VNA) is directly connected in transmission to the PCB, by means of a set of low-losses room temperature Cu and cryogenic BeCu coaxial cables (i.e. with a total signal attenuation of about $\text{\SI{-0.5}{\decibel}}$), delivering an excitation power ranging from $\text{\SI{-40}{\dBm}}$ up to $\text{\SI{-10}{\dBm}}$. Frequency scans are, then, run in order to locate the resonance frequency of each device, recording the complex $S_\text{21}$ transmission parameter.\\
Upon acquisition, the complex $S_\text{21}$ data are, subsequently, renormalised to the environmental factor $A(f) = a e^{i(\alpha-2\pi\tau f)}$, in order to remove any attenuation and phase shift introduced by the microwave cabling. The evolution of the different resonance signals, with an increasing excitation power, is plotted in \autoref{SUPPL_FIG_03_a}. In particular, for power levels larger than $\text{\SI{-36}{dBm}}$, it is possible to notice the typical resonance peak truncation associated to quasiparticles self-heating \cite{Thomas:2020}, superimposed to the asymmetric bending and red-shift characteristic of the self-Kerr non-linearity \cite{Anferov:2020, Kirsh:2021, Joshi:2022}. Interestingly, more pronounced deformations of the resonance signal are found to affect the devices characterised by a larger outermost nanotrench value, which might be interpreted as a first evidence of the effect of such a surface nanostructuring on the frequency response of the fabricated resonators.\\
Nevertheless, in order to avoid such non-linear effects to perturb the temperature monitoring performed through these nanostructured resonators, a suitable microwave excitation level of $\text{\SI{-40}{\dBm}}$ is selected for performing further cryogenic tests.

\subsection{Determination of the inductive coupling condition}
\label{SUPPL_MAT_03_b}

The influence of the coupling distance $d_\text{cpl}$ on the frequency response of the fabricated sensors is also investigated, by characterising the reference resonator at $\text{\SI{4.2}{\kelvin}}$ in liquid He. More in details, the device is excited, by means of a VNA delivering a $\text{\SI{-40}{\dBm}}$ during different LHe cool-downs, where the coupling distance in gradually increased, by changing the 3-D printed PLA support, from $\text{\SI{0.50}{\milli\metre}}$ up to $\text{\SI{3.00}{\milli\metre}}$.\\

\begin{table*}[!t]
\centering
\begin{threeparttable}
\caption{$f_{\text{res}}$, $Q_\text{L}$, $Q_\text{c}$, $Q_\text{i}$, $\phi$ fit parameters and $R^2$ goodness of fit coefficient, estimated on the reference resonator, at $\text{\SI{4.2}{\kelvin}}$ in LHe, for an increasing $d_\text{cpl}$. The omission of the related uncertainties, extracted through the $\SI{95}{\%}$ Confidence Intervals (CIs)s and always lower than $\SI{5}{\%}$, is intentional, for the sake of clarity and readability.}
\label{SUPPL_TABLE_03_b}
\renewcommand{\arraystretch}{1.25}
\begin{tabular*}{\textwidth}{@{\hspace{10pt}\extracolsep{\fill}} ccccccc @{\hspace{10pt}}}
\toprule
$d_{\text{cpl}}$ & $f_{\text{res}}$ & $Q_\text{L}$ & $Q_\text{c}$ & $Q_\text{i}$ & $\phi$  & $R^2$\\
(mm) & (GHz) & $\bigl(\times$10\textsuperscript{3}$\bigr)$ & $\bigl(\times$10\textsuperscript{6}$\bigr)$ & $\bigl(\times$10\textsuperscript{3}$\bigr)$ & (\textdegree) & (\%)\\
\midrule
0.50 & 1.3613 & 3.10 & 0.005 & 7.11 & 1.01 & 99.98\\
0.75 & 1.3640 & 5.25 & 0.011 & 8.85 & 1.43 & 99.98\\
1.00 & 1.3644 & 6.84 & 0.025 & 8.89 & 6.11 & 99.97\\
1.50 & 1.3640 & 8.39 & 0.099 & 9.04 & 9.89 & 99.81\\
2.00 & 1.3641 & 9.03 & 0.414 & 9.20 & 1.94 & 98.42\\
2.50 & 1.3639 & 9.47 & 0.904 & 9.55 & 8.21 & 96.23\\
3.00 & 1.3641 & 9.58 & 3.016 & 9.61 & 4.38 & 95.64\\
\bottomrule
\end{tabular*}
\end{threeparttable}
\end{table*}

The real and imaginary parts of the transmission $S_\text{21}$ data are first measured and, successively, fitted to a complex notch-type circular model \cite{Probst:2015}
\begin{equation}
\label{SUPPL_EQUATION_03_b_01}
S_{\text{21}}(f) = A(f)\:\biggl( 1-\frac{(Q_\text{L}/|Q_\text{c}|)e^{i\phi}}{1+2iQ_\text{L}(f/f_{\text{res}}-1)} \biggr)
\end{equation}
where the term 
\begin{equation}
\label{SUPPL_EQUATION_03_b_02}
A(f) = a e^{i(\alpha-2\pi\tau f)}
\end{equation}
is an environmental pre-factor, necessary to properly take into consideration the effect of the $a$ amplitude attenuation, $\alpha$ phase shift and $\tau$ time delay introduced by the microwave cabling. Additionally, $f_{\text{res}}$ stands for the resonance frequency of the device, while $\phi$ determines the impedance mismatches at the input/output ports. Finally, the finite microwave losses affecting the resonator are modelled by the loaded $Q_\text{L}$ and coupling $Q_\text{c}$ quality factors. The internal Q-factor $Q_\text{i}$ can also be estimated by recurring to the constitutive relation $Q_\text{L}^{-1} = Q_\text{i}^{-1} + Q_\text{c}^{-1}$ \cite{Pozar:2012}.\\

\begin{figure*}[t!]
\includegraphics[width=\textwidth]{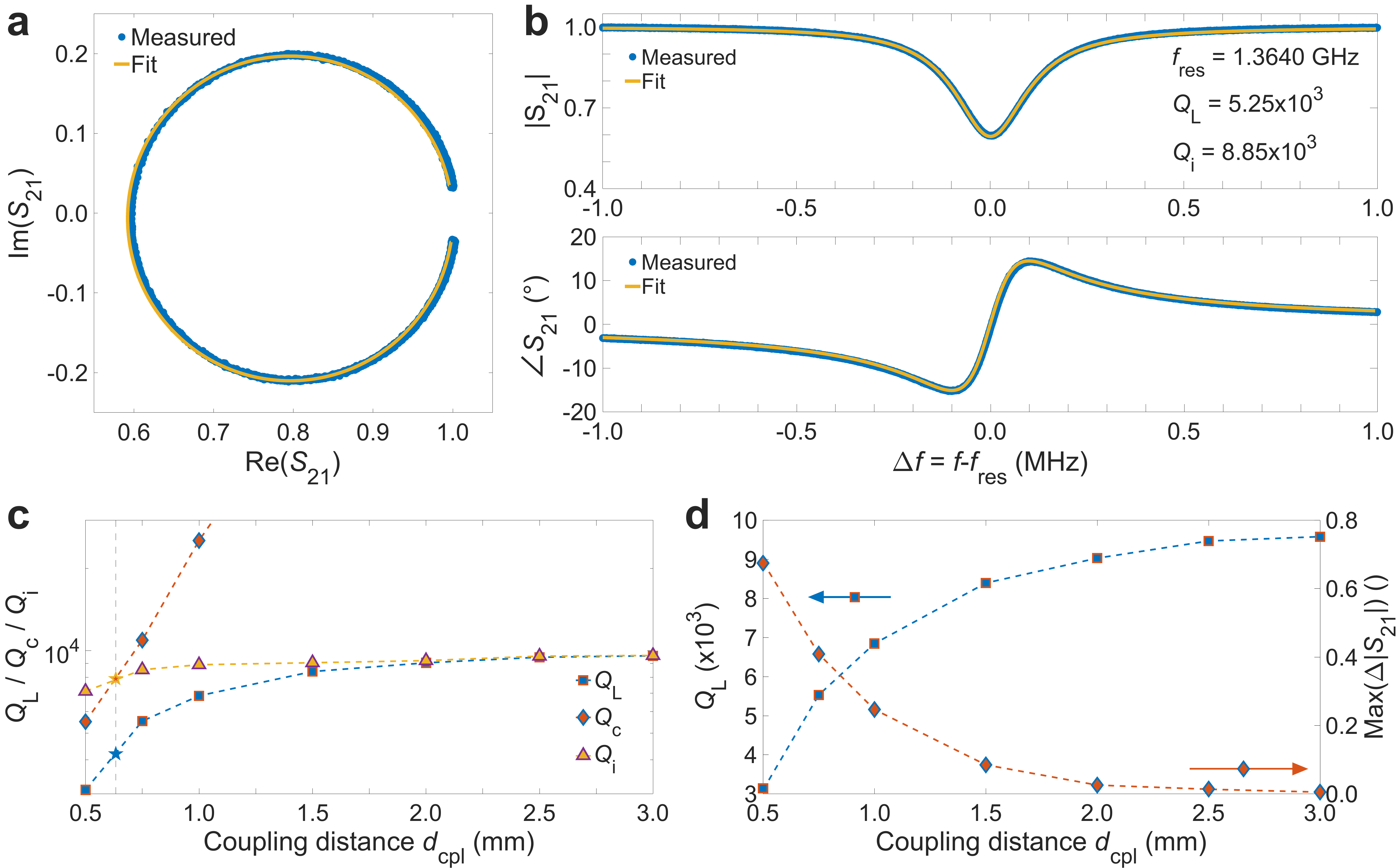}
\caption{Experimental determination of an optimal electromagnetic coupling condition, at $\text{\SI{4.2}{\kelvin}}$ in LHe. a) Circular plot of the complex $S_{\text{21}}$ parameter, for a $\text{\SI{0.75}{\milli\metre}}$ distance. A notch-type model (yellow line) is used to fit the experimental data (blue dots). b) $S_{\text{21}}$ magnitude and phase, for $d_{\text{cpl}} = \text{\SI{0.75}{\milli\metre}}$. c) Variation of $Q_\text{L}$, $Q_\text{c}$ and $Q_\text{i}$ with respect to $d_{\text{cpl}}$. The critical coupling condition is represented by the single pentagram points. d) Variation of $Q_\text{L}$ (blue square points) and resonance dip (orange diamond points).}
\label{SUPPL_FIG_03_b}
\makeatletter
\let\save@currentlabel\@currentlabel
\edef\@currentlabel{\save@currentlabel(a)}\label{SUPPL_FIG_03_b:a}
\edef\@currentlabel{\save@currentlabel(b)}\label{SUPPL_FIG_03_b:b}
\edef\@currentlabel{\save@currentlabel(c)}\label{SUPPL_FIG_03_b:c}
\edef\@currentlabel{\save@currentlabel(d)}\label{SUPPL_FIG_03_b:d}
\edef\@currentlabel{\save@currentlabel(a-b)}\label{SUPPL_FIG_03_b:a-b}
\edef\@currentlabel{\save@currentlabel(c-d)}\label{SUPPL_FIG_03_b:c-d}
\makeatother
\end{figure*}

In \autoref{SUPPL_TABLE_03_b} we report the resulting fitting parameters, including also the goodness of fit $R^2$ coefficient, found to be larger than $\text{\SI{95}{\percent}}$ for all the analysed coupling conditions. In particular, the accuracy of the fitting protocol is shown in \autoref{SUPPL_FIG_03_b:a-b}, where the coincidence between the measured resonance dip and the estimated expression from \autoref{SUPPL_EQUATION_03_b_01} is evident. As described by \autoref{SUPPL_FIG_03_b:c}, considering the evolution of the Q-factor for an increasing $d_\text{cpl}$ coupling distance, $Q_\text{L}$ is found to saturate to the unloaded condition for a $Q_\text{L}(\SI{3}{\milli\metre}) = \text{\SI{9.58}{}$\times$10\textsuperscript{3}}$ value. Additionally, the critical coupling, for which $Q_\text{i} = Q_\text{c}$ and, therefore, $Q_\text{L} = Q_\text{i}/2$, occurs at a distance of about $\text{\SI{0.60}{\milli\metre}}$.\\
As the resonance frequency dip shows an opposite trend to the evolution of the Q-factor, with an exponential decrease for larger coupling distances [see \autoref{SUPPL_FIG_03_b:d}], a slightly under-coupled value of $\text{\SI{0.75}{\milli\metre}}$ is selected for performing further cryogenic tests, in order to take benefit from the high Q-factor condition and a still large Signal-to-Noise Ratio (SNR).

\section{AFM scans of the different single-nanotrench 4-wire structures}
\label{SUPPL_MAT_04}

In this section, we report the Atomic Force Microscopy (AFM) scans of each single-nanotrench line, carried out with the intention to provide a morphological characterisation of such partially etched Nb\textsubscript{50}Ti\textsubscript{50} regions.\\

As fully described in the main text, the geometry of such a DC test structures consists of a $\text{\SI{100}{\mu\metre}}$ long resistive wire, of $\text{\SI{15}{\mu\metre}} \times \text{\SI{100}{\nano\metre}}$ rectangular cross-section, where a single partially etched nanotrench, extending all the across its width, is patterned at the centre of the current path, equally distanced from each lateral voltage reading line.\\
Once fabricated on the same wafer containing the different nanostructured resonators, AFM scans are exploited to inspect the quality of each single-nanotrench lines. In particular, the results corresponding to such a morphological characterisation are presented in \autoref{SUPPL_FIG_04}, where it is possible to extract the width of each feature, as well as the residual thickness of the partially etched thin-film. Through this analysis, a width deviation, from the design targeted values, lower than $\text{\SI{10}{\nano\metre}}$ is recorded for all the fabricated single nanotrenches, while a residual Nb\textsubscript{50}Ti\textsubscript{50} film thickness of about $\text{\SI{20}{\nano\metre}}$ is measured, consistent with the characterisation results performed on the nanostructured resonators and reported in main text.

\begin{figure*}[t!]
\centering
\includegraphics[width=0.568\textwidth]{suppl_fig_04.pdf}
\caption{AFM scans of the different single-nanotrench 4-wire DC test structures. a) Reference structure (no nanotrench). b) $\text{\SI{50}{\nano\metre}}$ wide nanotrench. c) $\text{\SI{110}{\nano\metre}}$ wide nanotrench. d) $\text{\SI{170}{\nano\metre}}$ wide nanotrench. e) $\text{\SI{230}{\nano\metre}}$ wide nanotrench. f) $\text{\SI{290}{\nano\metre}}$ wide nanotrench. g) $\text{\SI{350}{\nano\metre}}$ wide nanotrench.}
\label{SUPPL_FIG_04}
\makeatletter
\let\save@currentlabel\@currentlabel
\edef\@currentlabel{\save@currentlabel(a)}\label{fig_03_09:a}
\edef\@currentlabel{\save@currentlabel(b)}\label{fig_03_09:b}
\edef\@currentlabel{\save@currentlabel(c)}\label{fig_03_09:c}
\edef\@currentlabel{\save@currentlabel(d)}\label{fig_03_09:d}
\edef\@currentlabel{\save@currentlabel(e)}\label{fig_03_09:e}
\edef\@currentlabel{\save@currentlabel(f)}\label{fig_03_09:f}
\edef\@currentlabel{\save@currentlabel(g)}\label{fig_03_09:g}
\makeatother
\end{figure*}

\section{Typical $I$/$V$ dataset for a sweep of the cryogenic temperature $T$}
\label{SUPPL_MAT_05}

\begin{figure*}[t!]
\includegraphics[width=\textwidth]{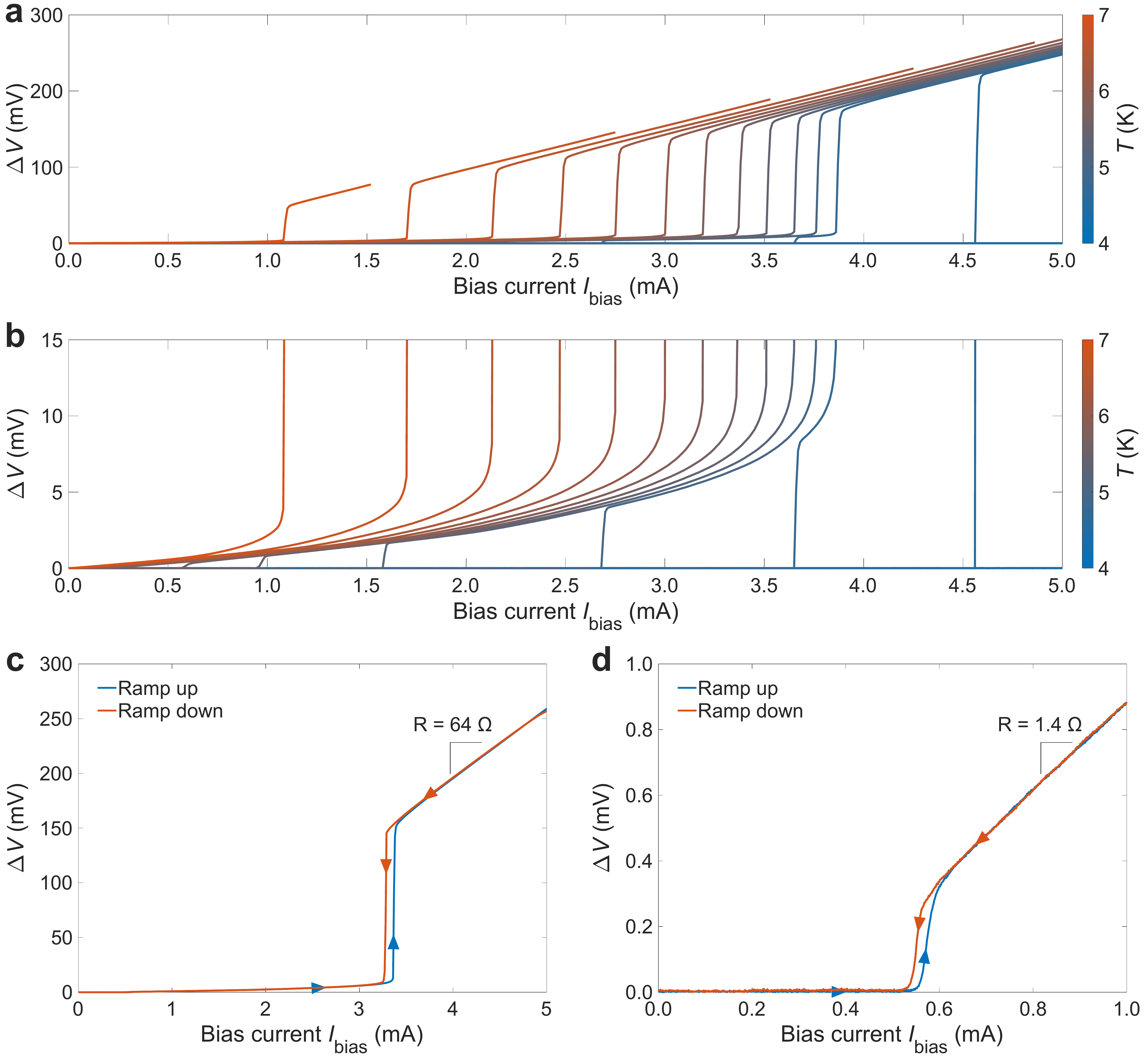}
\caption{Effect of the temperature on the superconducting critical current transition. a) Set of I/V scans for the $\text{\SI{224}{\nano\metre}}$ single-nanotrench 4-wire structure, at different cryogenic temperatures. b) Zoom on a $\text{\SI{0}{}}-\text{\SI{15}{\milli\volt}}$ interval, to highlight the presence of two distinct superconducting transitions. c) Higher-current transition at $\text{\SI{5.6}{\kelvin}}$, associated to the whole superconducting 4-wire device. d) Lower-current transition at $\text{\SI{5.6}{\kelvin}}$, associated to the partially etched feature.}
\label{SUPPL_FIG_05}
\makeatletter
\let\save@currentlabel\@currentlabel
\edef\@currentlabel{\save@currentlabel(a)}\label{fig_03_10b:a}
\edef\@currentlabel{\save@currentlabel(b)}\label{fig_03_10b:b}
\edef\@currentlabel{\save@currentlabel(c)}\label{fig_03_10b:c}
\edef\@currentlabel{\save@currentlabel(d)}\label{fig_03_10b:d}
\makeatother
\end{figure*}

The aim of this section is to report one typical dataset dataset of $I$/$V$ curves, for the DC characterisation of the superconducting transition of a single-nanotrench structure (i.e. the $\text{\SI{224}{\nano\metre}}$ wide one). In this perspective, as also described in the main manuscript, the extraction of the DC critical current density $J_\text{C}(T)$ is performed by recurring to a fixed voltage threshold of about $\SI{1}{\mu\volt}$, corresponding to the RMS voltage noise level of the characterisation setup. More specifically, considering the $\SI{100}{\mu\metre}$ separation between the two voltage probes, this threshold determines an average electric field criterion expressed as $E_\text{th}=\SI{100}{\mu\volt/\centi\metre}$.\\

When performing such a characterisation, with variable current and temperature, two distinguishable transitions are always recorded, with the first one being associated to the partially etched region, while the second one represents the collapsing of the overall superconducting film. In such a sense, \autoref{SUPPL_FIG_05} reports the typical set of $I$/$V$ scans obtained at different temperatures (in this case, for the $\text{\SI{224}{\nano\metre}}$ wide nanotrench). In particular, the resistance values, extracted from a linear interpolation of the $I$/$V$ curves above the two $I_\text{C}$, turn out to be about $\text{\SI{64}{\ohm}}$ and $\text{\SI{1.4}{\ohm}}$, respectively for the first and second transition. Such results are consistent with the values that may be derived from the estimation of the Nb\textsubscript{50}Ti\textsubscript{50} sheet resistance at different thicknesses \cite{Chatel:2025}, corresponding to $\text{\SI{62}{\ohm}}$, for the whole $\text{\SI{100}{\nano\metre}}$ thick 4-wire section, and $\text{\SI{1.1}{\ohm}}$ for the $\text{\SI{20}{\nano\metre}}$ thick and $\text{\SI{230}{\nano\metre}}$ wide nanotrench.\\
Moreover, even though the first transition occurs at lower current values, as expected, for an increasing temperature, also the second one seems to follow the same trend. We suspect that such a behaviour can be ascribed to a thermal effect, induced by the heating of the thinned Nb\textsubscript{50}Ti\textsubscript{50} region once transited, which might eventually increase the temperature of the device and make the transition of the overall 4-wire structure to occur at lower current values.

\printbibliography[heading=bibintoc]